\documentclass[twocolumn]{aastex631}
\usepackage[T1]{fontenc}
\usepackage{amsmath}
\usepackage{amssymb}
\usepackage{soul}
\usepackage{graphicx} % Required for inserting images
\usepackage{subfigure}
\begin{document}

\title{Could the neutrino emission of TXS~0506+056 come from the accretion flow of the supermassive black hole?}

\author[0009-0003-7748-3733]{Qi-Rui Yang}
\affiliation{School of Astronomy and Space Science, Nanjing University, Nanjing 210023, China; xywang@nju.edu.cn}
\affiliation{Key laboratory of Modern Astronomy and Astrophysics (Nanjing University), Ministry of Education, Nanjing 210023, China}

\author[0000-0003-1576-0961]{Ruo-Yu Liu}
\affiliation{School of Astronomy and Space Science, Nanjing University, Nanjing 210023, China; xywang@nju.edu.cn}
\affiliation{Key laboratory of Modern Astronomy and Astrophysics (Nanjing University), Ministry of Education, Nanjing 210023, China}
\affiliation{Tianfu Cosmic Ray Research Center, Chengdu 610000, Sichuan, China}

\author[0000-0002-5881-335X]{Xiang-Yu Wang}
\affiliation{School of Astronomy and Space Science, Nanjing University, Nanjing 210023, China; xywang@nju.edu.cn}
\affiliation{Key laboratory of Modern Astronomy and Astrophysics (Nanjing University), Ministry of Education, Nanjing 210023, China}
%\correspondingauthor{Xiang-Yu Wang}
%\email{xywang@nju.edu.cn}

\begin{abstract}
High-energy neutrinos  from the blazar TXS~0506+056 are usually thought to arise from the relativistic jet pointing to us. However, the composition of jets of active galactic nuclei (AGNs), whether they are baryon dominated or Poynting flux dominated, is largely unknown. In the latter case, no comic rays and neutrinos would be expected from the AGN jets. In this work, we study whether the neutrino emission from TXS~0506+056 could be powered by the accretion flow of the supermassive black hole. 
%We propose that cosmic rays could be accelerated in the magnetical arrested disk (MAD) surrounding the black hole and neutrinos are produced in the central region through $pp$ and $p\gamma$ processes. 
Protons could be accelerated by magnetic reconnection or turbulence in the inner accretion flow. To explain the neutrino flare of TXS~0506+056 in the year of 2014-2015, a super-Eddington accretion  is needed. During the steady state, a sub-Eddington accretion  flow could power 
a steady neutrino emission that may explain the long-term  neutrino flux from TXS 0506+056.  We consider the neutrino production in both magnetically arrested accretion (MAD) flow and the standard
and normal evolution (SANE) regime of accretion. In the MAD scenario,   due to a high magnetic field, a large dissipation radius  is required to avoid the cooling of protons and secondary pions. 
%To suppress the cooling of high-energy protons due to synchrotron emission in  the magnetic field, the plasma $\beta$ should satisfy $\beta>1$ or the acceleration region has a large radius.  

\end{abstract}

\section{Introduction}
Active Galactic Nuclei (AGN) are prime candidate sources of the high-energy astrophysical neutrinos.
In 2017, IceCube detected a high-energy neutrino
event in the direction coincident with the blazar TXS~0506+056, which was found to be flaring at the gamma-ray band
\citep{Aartsen2018Sci...361.1378I}. A follow-up analysis of archival IceCube neutrino data revealed an earlier outburst of neutrinos from the same source in 2014/2015 without
an accompanying flare of gamma rays\citep{Aartsen2018Sci...361..147I}. Later, an independent search for point-like sources in the northern hemisphere using ten years of IceCube data revealed that
TXS~0506+056 is coincident with the second hottest-spot of the neutrino event excess \citep{2022Sci...378..538I}. Since blazars are AGNs with relativistic jets pointing toward our line of sight, the Doppler effect remarkably boosts the flux of blazars received by us. Most of the previous studies ascribe the high-energy neutrino emission of TXS~0506+056 to relativistic protons accelerated in the jet, through either photopion production with the radiation of the jet itself and the external radiation of the surrounding environment \cite[e.g.][]{Murase_2018, Keivani2018, Gao2019NatAs...3...88G, Cerruti2019, Rodrigues2019, Xue2019, ZhangB2020, Xue2021ApJ...906...51X} , or proton-proton collisions with matter of the jet and cloud/star entering the jet \citep[e.g.][]{Sahakyan2018, Liu2019, Banik2020, Wang2022}.  { However, the composition of jets of AGNs, whether they are baryon dominated or Poynting flux dominated, is largely unknown. The composition of jets
is tightly related to their formation mechanisms.  As discussed by
\citet{2001bhbg.conf..206C}, electromagnetically dominated
outflows, such as those generated by the extraction of the spin energy of the black hole, are pair-dominated jets with a low baryonic pollution.  In this case, no comic rays and neutrinos would be expected from the AGN jets. }

{\citet{2022Sci...378..538I} reported an excess of neutrino
events associated with NGC 1068, a nearby type-2 Seyfert galaxy , with a significance of 4.2$\sigma$. 
Seyfert galaxies are radio-quiet AGNs with much weaker jets compared to blazars. In NGC 1068, a supermassive black hole (SMBH) at the center  is  highly obscured by thick gas and dust \citep{2022Natur.602..403G}. X-ray studies have suggested that NGC 1068 is among the
brightest AGNs in intrinsic X-rays \citep{2015ApJ...812..116B}, which  is generated through Comptonization of accretion-disk photons in hot
plasma above the accretion disk, namely the coronae. Given the dense matter and the intense radiation in the environment, efficient neutrino  production  is  expected if cosmic rays are accelerated at the proximity of the SMBH \citep{Murase2020PhRvL.125a1101M, 2020ApJ...891L..33I}.
Interestingly, the reported neutrino flux is higher than the GeV gamma-ray flux, implying that gamma-rays above 100\,MeV are strongly attenuated by  
dense X-ray photons while neutrinos can escape.} 

A correlation between unabsorbed hard X-rays and neutrinos in radio-loud and
radio-quiet AGN is suggested by \citet{Kun2024arXiv240406867K},  raising the possibility of a common neutrino production
mechanism involved in both types of AGNs. Since neutrino emission from radio-quiet Seyfert galaxies is unlikely related to their weak
jets, a natural question arises as to whether neutrinos
from blazars can be produced somewhere besides
their powerful jets, such as from the proximity of the SMBH.  
%In addition, the maximum likelihood stacking searches for cumulative neutrino flux from the second Fermi-LAT AGN catalog (2LAC) as well as the point-source searches using the IceCube muon track events and blazars in Fermi-LAT 3LAC have independently shown that Fermi-LAT-resolved blazars only contribute a small portion of the IceCube cumulative neutrino flux (Aartsen et al. 2017a; Pinat \& Sánchez 2017; Hooper et al.2019).
%Abbasi et al. (2024) performed an in-depth stacking study of the correlation between the radio-loud AGN and ten years of neutrino data from IceCube. No significant correlation was found.  When compared to the IceCube diffuse flux, at 100 TeV and for a spectral index of 2.5, the upper limits derived are ~3\% ~9\%. This implies that the radio loudness does not play a key role in the neutrino production.

Hadronic interactions responsible for the neutrino production also generate gamma rays with comparable flux and energy spectra to that of neutrinos. 
If such interactions take place near the SMBH,
where intense infrared-optical photons from the accretion disk and X-rays from the hot corona are present, pair production and subsequent electromagnetic cascade will reprocess the gamma rays into keV-MeV photons. The apparent underproduction of gamma-rays compared to neutrinos
in Seyfert galaxies is naturally explained in such environments.  A hint of strong gamma-absorption in neutrino sources is also
observed in the diffuse neutrino flux \citep{Murase2020PhRvL.125a1101M}, pointing to the so-called "hidden"  neutrino sources.

Motivated by the above reasonings, in this paper, we investigate the possibility that the neutrino outburst of TXS~0506+056 and the steady neutrino emission are produced by the core region of the AGN. Indeed, there have been suggestions that particles may be accelerated in the accretion disk via magnetic reconnection and turbulence \citep[e.g.][]{Yuan2003, Pino2010, Hoshino2013, Kunz2016, Ripperda2022,Kheirandish2021ApJ...922...45K}. If the SMBH has a fast-rotating magnetosphere, the centrifugal force may also serve as an efficient particle accelerator\citep{Gangadhara97, Rieger2008}.

The rest part of the paper is organized as follows. We introduce our model in Section 2 and show the results in Section 3. The discussions are given in Section 4. Finally we summarize the results in Section 5.

\section{Neutrinos from the AGN disk in TXS 0506+056 ? }
The blazar TXS 0506+056 is the first individual neutrino source identified at $ > 3\sigma$ significance excess and $13 \pm 5$ high-energy neutrino events was discovered in the period between September 2014 and March 2015. \citep{Aartsen2018Sci...361..147I}. 
For this flare period, the neutrino integrated luminosity (per flavor) between 32 TeV and 4 PeV  is estimated to be $L_{\nu_{\mu}} \sim  10^{47} \rm{erg~s^{-1}}$. The time-integrated analysis  of the ten years of IceCube data revealed that
TXS~0506+056 is coincident with the second hottest-spot of the neutrino event excess \citep{2022Sci...378..538I}. A mean flux of $\sim 10^{-13}\,{\rm TeV cm^{-2} s^{-1}}$ is obtained \citep{2022Sci...378..538I}, corresponding to  a neutrino  luminosity of $L_{\nu_{\mu}} \sim 5\times 10^{44} \rm{erg~s^{-1}}$.

%The blazar TXS 0506+056 is the first individual neutrino source identified at $ > 3\sigma$ significance. (IceCube Collaboration 2018). 
%For TXS 0506+056, the neutrino integrated luminosity between 40 TeV and 4 PeV calculates as $L_{\nu_{\mu}+\bar{\nu}_{\mu}} = 5.6 \times 10^{44} \rm{erg~s^{-1}}$, with corresponding integrated lower and upper limits of $7.8 \times 10^{43} \rm {erg~s^{-1}}$ and $1.3 \times 10^{45} \rm {erg~s^{-1}}$, respectively. (Abbasi et al. 2022)
The Eddington luminosity of the accretion disk in TXS 0506+056 is estimated to be
\begin{equation}
    L_{\rm Edd}=4\times 10^{46} {\rm erg~s^{-1}} \left(\frac{M_{\rm BH}}{3\times10^8 M_\odot}\right),
\end{equation}
where $M_{\rm BH}$ is the mass of the supermassive black hole. The Eddington accretion rate is defined as $\dot{M}_{\rm Edd}=L_{\rm Edd}/({\eta_a}c^2)$, where $\eta_a=0.1$ \citep{Yuan2014ARA&A..52..529Y}  Assuming an efficiency $\epsilon_{\rm CR}$ for accretion power converted into cosmic rays, the cosmic ray luminosity is estimated to be
\begin{equation}
\begin{aligned}
    L_{\rm CR}&=\epsilon_{\rm CR} \dot{M} c^2\\
    &= 4\times 10^{46} {\rm erg~s^{-1}}\left(\frac{\epsilon_{\rm CR}}{0.1}\right)  \left(\frac{\dot{m}}{1}\right) \left(\frac{M_{\rm BH}}{3\times10^8 M_\odot}\right), 
\end{aligned}
\end{equation}
where $\dot{m} = \dot{M}/\dot{M}_{\rm Edd}$ is the dimensionless accretion rate and we use $\epsilon_{\rm CR}= 0.1$  as a fiducial value. 
%We assume the injection fraction of $\epsilon_{\rm CR}\sim 0.33$ can be converted to  
Then the neutrino luminosity (per flavor) produced by cosmic rays via $pp$ or $p\gamma$ interaction is given by
\begin{equation}
\begin{aligned}
    L_{\rm \nu_\mu}&=\frac{1}{8}f_{pp,p\gamma}L_{CR}\\
    &\sim 5\times10^{45}  {\rm erg~s^{-1}}f_{pp,p\gamma}\left(\frac{\epsilon_{\rm CR}}{0.1}\right)  \left(\frac{\dot{m}}{1}\right) \left(\frac{M_{\rm BH}}{3\times10^8 M_\odot}\right).
\end{aligned}
\end{equation}
Here $f_{pp,p\gamma} = t_{\rm loss}/t_{pp,p\gamma}$ is the efficiency of $pp$ and $p\gamma$ processes expressed in the ratio between the proton energy loss timescale ($t_{\rm loss}$) and the hadronic interaction timescale ($t_{pp,p\gamma}$), which will be calculated in following sections. To explain the observed neutrino luminosity, we require 
\begin{equation}
    \dot {m} \sim 20 f_{pp,p\gamma}^{-1}  \left(\frac{\epsilon_{\rm CR}}{0.1}\right)^{-1}  \left(\frac{M_{\rm BH}}{3\times10^8 M_\odot}\right)^{-1} \left(\frac{L_{\nu_{\mu}}}{10^{47}{\rm erg s^{-1}}}\right)^{-1}.
\end{equation}
Therefore, to explain the neutrino flare during 2014-2015, a  super-Eddington accretion rate is required. On the other hand, to explain the 10-year quasi-steady-state neutrino emission with a flux  2 orders of magnitude lower, a sub-Eddington accretion may be viable. 

%\subsection{Neutrino production efficiency for $pp$ and $p\gamma$ interactions}
\subsection{The process of proton acceleration and cooling in the accretion flow}
High-energy protons may be accelerated  by magnetic reconnection, stochastic acceleration via MHD turbulence or electric potential gaps in the black hole magnetosphere. We here do not specify the detailed acceleration mechanism, but phenomenologically parameterize the particle acceleration timescale by
\begin{equation}
    t_{\rm acc} \approx \frac{\eta r_{\rm L}}{c},
\end{equation}
where the different particle acceleration mechanisms may be characterized by distinct parameter $\eta$, which could be understood as the particle acceleration efficiency. $r_{\rm L} = E/eB$ is the Larmor radius. The maximum energy of non-thermal proton is determined by the balance among particle acceleration, cooling, and escape processes in the accretion flow. The escape term is common for all components. We consider diffusion and advection (infall to the BH) as the escape processes, whose timescales are estimated to be $t_{\rm diff} \approx R^2/D_{\rm R}$ and $t_{\rm fall} \approx R/V_{\rm R}$ respectively, where $D_{\rm R} = D_{\rm R}\approx \eta r_{\rm L} c/3$ is the diffusion coefficient and $V_{\rm R}$ is the radial velocity of the accretion flow. The total escape time is given by $t_{\rm esc}^{-1} = t_{\rm diff}^{-1}+t_{\rm fall}^{-1}$. 

For the cooling of cosmic ray protons, we consider inelastic collisions ($pp$), photomeson production ($p\gamma$), Bethe-Heitler pair production, and proton synchrotron radiation.
%For TXS 0506+056, the core origin of hard X-ray was argued possible, since the different light curve between $\gamma$-ray attributed to jet component and hard X-ray emission \citep{Kun2024arXiv240406867K}. 
%The "big blue bump", which has been seen in many AGN, is considered to be the photon field of TXS 0506+056. 
%The "big blue bump" is attributed to multi-temperature blackbody emission from a geometrically thin, optically thick disk and hard X-ray from comptonlized corona. 
The photon field includes the multi-temperature black-body emission from the accretion disk and hard X-ray emission from comptonized corona.
The coronal spectrum can be modeled by a power law with an exponential cutoff. The photon index of TXS 0506+056, $\Gamma_{X}$, varies between 1.5 – 1.9 among observations \citep{Acciari2022ApJ...927..197A}. The photon index is correlated with $\lambda_{\rm Edd} = L_{\rm bol}/L_{\rm Edd} \sim 0.24$ as $\Gamma_{X} \approx 0.167 \times \rm{log}(\lambda_{\rm Edd}) + 2.0 = 1.89$ \citep{Padovani2019MNRAS.484L.104P,Trakhtenbrot2019NatAs...3..242T}, and the cutoff energy is given by $\varepsilon_{\rm X,cut} \sim -74 \rm{log}(\lambda_{Edd}) + 1.5 \times 10 keV = 0.19MeV$ \citep{Ricci2018MNRAS.480.1819R}. We use the average hard X-ray  luminosity in 15–55 keV of $ (9.0 \pm 2.4) \times 10^{44}~\rm{erg~s^{-1}}$ to normalize the quasi-steady state X-ray component  \citep{Kun2024arXiv240406867K}. In the flare state characterized by an accretion rate two orders of magnitude higher, the bolometric luminosity would increase by a factor of 5-6 compared with  that in the steady state \citep{Huang2020ApJ...895..114H}.
%We assume that the corona region is confined within $R = \mathcal{R} R_g$, where $R_g = GM_{\rm BH}/c^2$ represents the gravitational radius of the SMBH. 

For the radiation of the disk, we consider a multi-temperature blackbody emission with the maximum temperature near the central supermassive black hole $T_{\text{disk}} \approx 0.49 (G M_{\rm BH} \dot{M}/(72 \pi \sigma_{\text{SB}} R_S^3))^{1/4} K$ \citep{Pringle1981ARA&A..19..137P}. The temperature of the disk can be expressed as $T(r) \approx (r/R_S)^{-3/4}$. Here, $M_{\rm BH}$ is the SMBH mass,  $R_S = 2GM_{\rm BH}/c^2$ is the Schwarzschild radius, and $\sigma_{\text{SB}}$ is the Stefan-Boltzmann constant. We can calculate the disk luminosity as 
\begin{equation}
    L_{\nu} = \frac{8\pi^2 h\nu^3}{c^2}\int_{R_S}^{R}\frac{rdr}{e^{(h\nu/kT(r))}-1}.
    \label{mbb}
\end{equation}
%Given that high-energy protons has the shortest timescale interacting with low energy photons, we take into $n_{\gamma} = L_{\gamma}/4\pi R^2c\varepsilon_0 \approx 1.6\times 10^{13}~\rm{cm}^{-3}(\textit{L}_{\gamma}/1.3 \times 10^{44}~erg~s^{-1})(\varepsilon_0/2~eV)^{-1} (\textit{R}/60\textit{R}_g)^{-2} $ is the number density of photons of inner disk.

The timescale of photomeson process ($p\gamma$) is $t_{p\gamma}\approx 1/(n_{\gamma}\sigma_{p\gamma}\kappa_{p\gamma}c)$,
where ${\sigma}_{p\gamma} \approx 5\times 10^{-28}~\rm{cm}^{2}$ is the  cross section for the photomeson process and $\kappa_{p\gamma} \sim 0.2$ is the inelasticity for $p\gamma$. The Bethe-Heitler energy loss rate is $t_{\rm B-H}\approx 1/(n_{\gamma}\hat{\sigma}_{\rm B-H}c)$, where $\hat{\sigma}_{\rm B-H}\sim 0.8\times10^{-30} \rm{cm}^{-3}$ is the effective cross section for the Bethe-Heitler process \citep{Murase2020PhRvL.125a1101M}.
The $pp$ cooling timescale is  $t_{pp} \approx 1/(n_{p}{\sigma}_{pp}\kappa_{pp}c)$, where ${\sigma}_{pp} \simeq 4\times 10^{-26} \rm{cm}^{2}$ and $\kappa_{pp} \approx 0.5$ are cross section and inelasticity for $pp$ process \citep{Kelner2006PhRvD..74c4018K}. The proton synchrotron timescale is  $t_{p,\rm{syn}} = {6 \pi m_p c}/({\gamma_p \sigma_T B^2})$. 
The total cooling rate can be given by $t_{\rm cool}^{-1} = t_{p\gamma}^{-1}+t_{pp}^{-1}+t_{\rm syn}^{-1}+t_{\rm B-H}^{-1}$, which is summation of all cooling rate. For high energy protons, the total energy loss rate is $t_{\rm loss}^{-1} = t_{\rm esc}^{-1} + t_{\rm cool}^{-1}$.
%\subsection{Neutrino Spectrum from the AGN core in TXS 0506+056}

\subsection{Proton spectrum}
As will be shown later, cooling timescale of high-energy protons is much shorter than the 14-15 outburst duration, so that we may consider a quasi-steady state spectrum for protons. To obtain the non-thermal spectra for protons, we solve the transport equation
\begin{equation}
    \frac{d}{dE_p} \left( -\frac{E_p}{t_{\text{cool}}} N_{p} \right) = \dot{N}_{\textit{p},\text{inj}} - \frac{N_p}{t_{\text{esc}}},
\end{equation}
where $N_{p} = dN/dE_p$, and $\dot{N}_{p,\text{inj}}$ is the injection function. We consider the injection  as a power-law distribution function with an exponential cutoff
\begin{equation}
    \dot{N}_{\textit{p},\text{inj}} = \dot{N}_0E_p^{-s_{\rm inj}}\exp\left(-\frac{ \textit E_{{p}}}{\textit E_{\textit{p},\text{max}}}\right),
\end{equation}
which is normalized by 
\begin{equation}
    \int E_p \dot{N}_{E_p,\text{inj}} dE_p = L_{\rm CR}.
\end{equation}
Here we consider the injected protron with spectrum index of $s_{\rm inj} = 2$. The maximum energy $E_{\textit{p},\text{max}}$  will be discussed in the following sections.
Then, the steady energy distribution of protons 
can be obtained  approximately by solving the following transport equation, 
\begin{equation}
\begin{aligned}
     \frac{dN_p}{dE_p} & = \frac{t_{\rm cool}}{E_p}\int^{\infty}_{E_p} dE\dot{N}_{E,\text{inj}}~\rm{exp}\left(-\int^{\textit E}_{\textit E_p}\frac{\textit t_{\text{cool}} \textit d\varepsilon_p}{\textit t_{\rm esc} \varepsilon_p}\right)\\
     & \approx \dot{N}_{E_p,\text{inj}}~t_{\rm loss} .
\end{aligned}
\end{equation}
%Therefore, we can derive the neutrino spectrum via $pp$ and $p\gamma$ process by using 
%\cite{Kelner2006PhRvD..74c4018K} and \cite{K&A2008PhRvD..78c4013K}. Here we consider the injected protron with spectrum index of $s_{\rm inj} = 2$. The maximum energy will be discussed in the following sections.

\begin{figure}
\centering
\subfigure[]{
\includegraphics[width=0.45\textwidth]{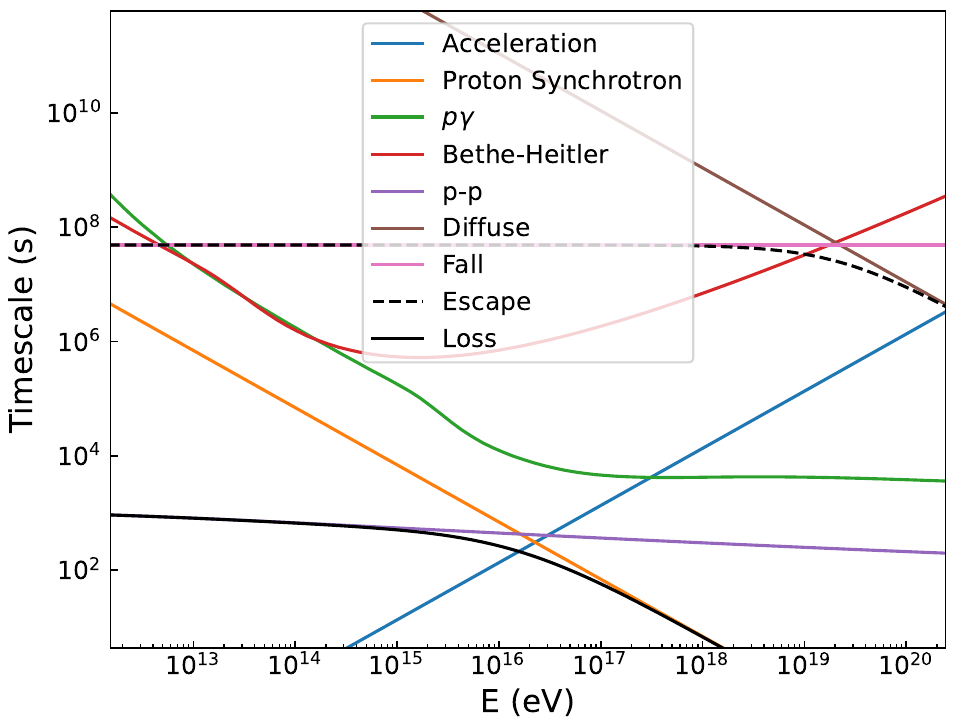}}
\subfigure[]{
\includegraphics[width=0.45\textwidth]{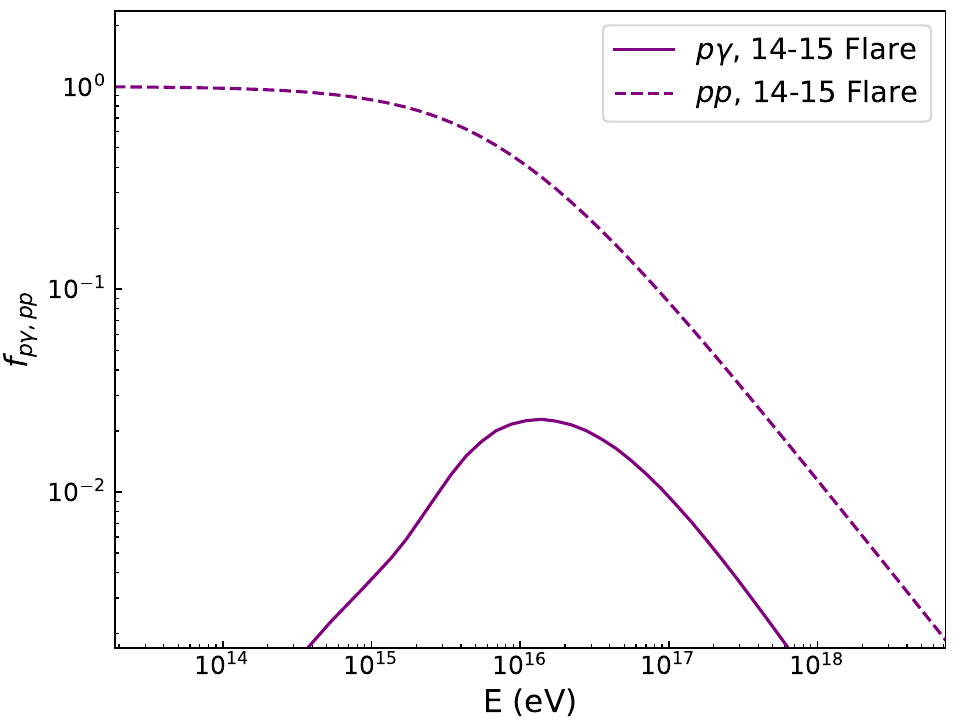}}
\subfigure[]{
\includegraphics[width=0.45\textwidth]{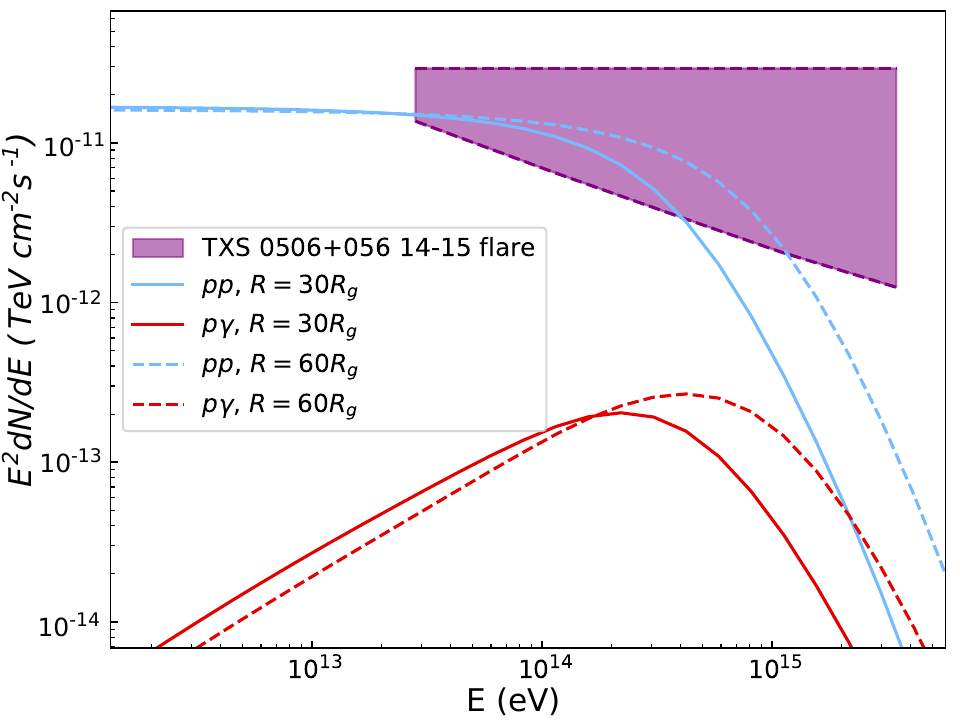}}
\caption{The panels, from top to bottom, show the various timescales, the efficiencies of $pp$ and $p\gamma$ interactions, and the neutrino spectrum in the MAD scenario. In panel (c), the observed neutrino spectrum of TXS 0506+056 during the 2014-2015 neutrino flare is also shown \citep{Aartsen2018Sci...361..147I}. We adopt the parameter values of $M_{\rm BH} = 3\times10^8 M_{\odot}$, $\epsilon_{\rm CR} = 0.1$,  $\epsilon = 0.01$, $\eta=300$ and $\dot{m}= 10$. In panel (a) and panel (b), the dissipation radius is set as $30R_g$, while in panel (c), two radius with $R =30R_g$ and $R =60 R_g$ are assumed. }
\label{fig:MAD Flare}
\end{figure}

\begin{figure}
\centering
\subfigure[]{
\includegraphics[width=0.45\textwidth]{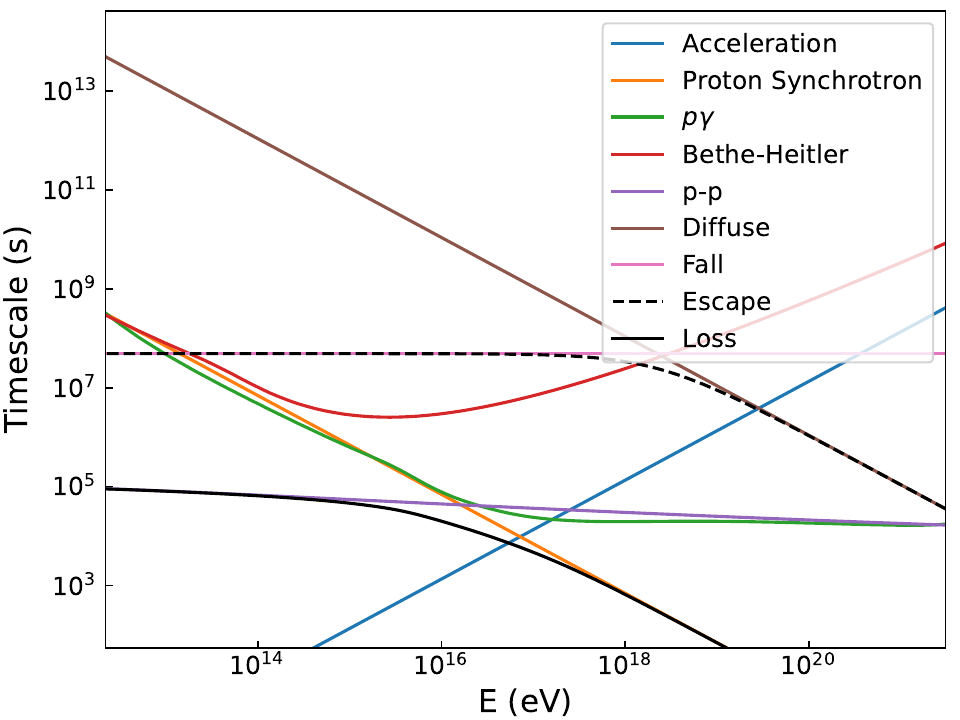}}
\subfigure[]{
\includegraphics[width=0.45\textwidth]{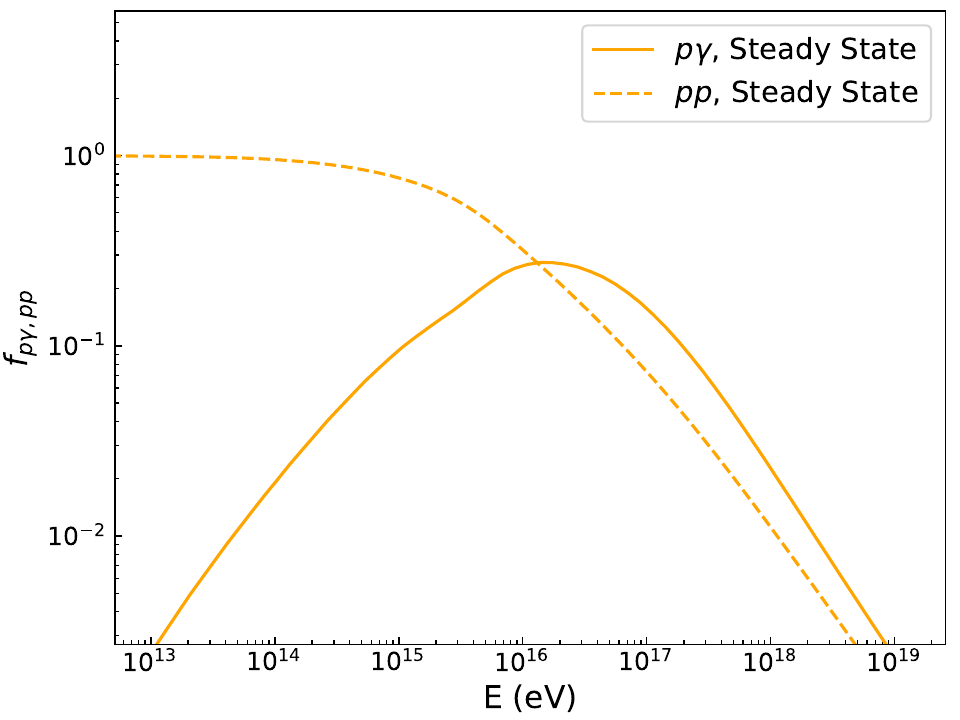}}
\subfigure[]{
\includegraphics[width=0.45\textwidth]{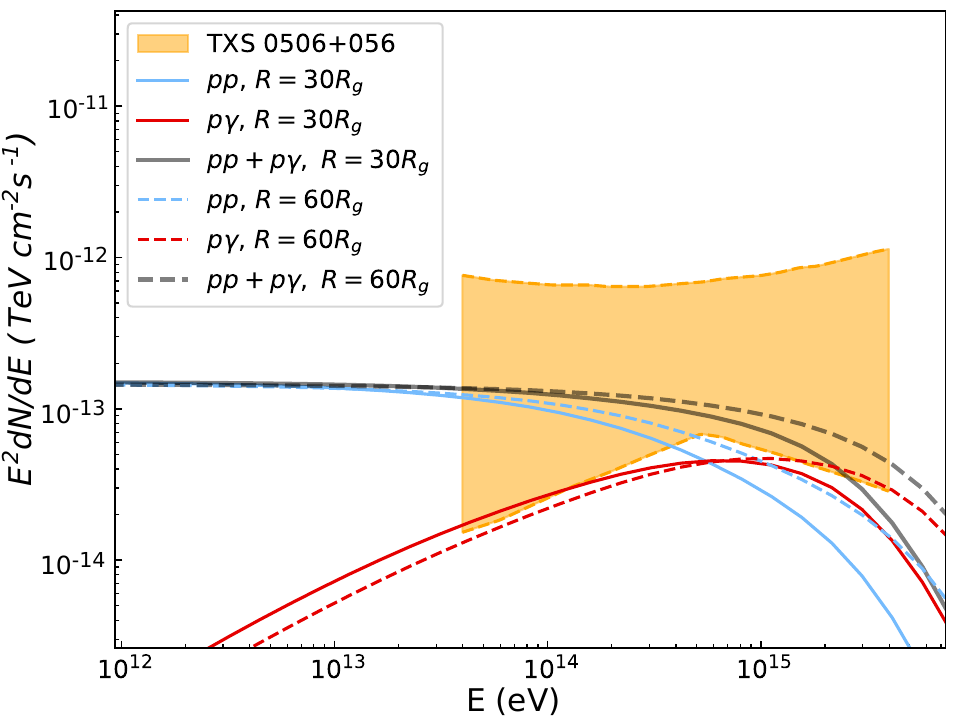}}
\caption{Same as Fig.~\ref{fig:MAD Flare}, but assuming $\dot{m} = 0.1$ to explain the  time-integrated  neutrino emission of TXS 0506+056 \citep{2022Sci...378..538I}.}
\label{fig:MAD Steady}
\end{figure}

\section{Neutrino emission from the accretion flow in the MAD and SANE scenarios}
A magnetically arrested accretion disc (MAD, see \cite{NIA2003PASJ...55L..69N,Bisnovatyi1974Ap&SS..28...45B,Igumenshchev2003ApJ...592.1042I,Tchekhovskoy2014MNRAS.437.2744T})
may be present in TXS~0506+056 since magnetohydrodynamical (MHD) simulations have shown that it can launch powerful
jets (e.g. \citet{Tchekhovskoy2011MNRAS.418L..79T}).
%TXS~0506+056 likely hosts a magnetically arrested disk (MAD) because it  efficiently launch relativistic jets by the Blandford-Znajek mechanism \citep{Tchekhovskoy2011MNRAS.418L..79T,Chael2019MNRAS.486.2873C,EHT2019ApJ...875L...5E,Porth2019ApJS..243...26P}. 
In this situation,  a large-scale poloidal magnetic field prevents gas from
accreting continuously at a magnetospheric radius. Around the magnetospheric radius, the
gas flow breaks up into a blob-like stream and moves inward by diffusing via magnetic interchanges through the magnetic field. 
We also consider the  standard
and normal evolution (SANE) regime of accretion, where the magnetic field that accumulates
around the BH is relatively weak.

%The magnetic field strength in the  accretion flows depends on the configuration of the magnetic field: $\beta \lesssim 10$ for MAD (magnetically arrested disk), whereas $\beta \gtrsim 10$ for SANE (standard and normal evolution). 

%The neutrinos produced by pion decay carry $1/8$ of the energy lost by protons to pion production, since charged and neutral pions are produced with roughly equal probability and muon neutrinos carry roughly $1/4$ of the pion energy in pion decay. In the neutrino energy range of interest (40\,TeV - 4\,PeV), the corresponding proton energy is in the range of 320\,TeV to 32\,PeV. The efficiency can be calculated as $f_{pp,p\gamma} \approx t_{pp,p\gamma}^{-1}/t_{\rm loss}^{-1}$.
\subsection{The MAD scenario}
MADs dissipate their magnetic energies through plasma processes,
such as magnetic reconnection \citep{Ball2018ApJ...853..184B,Ripperda2020ApJ...900..100R}, and nonthermal particles are efficiently accelerated by
reconnection \citep{Hoshino2012PhRvL.108m5003H,Sironi2014ApJ...783L..21S,Werner2018MNRAS.473.4840W} and/or turbulence \citep{Lynn2014ApJ...791...71L,Kimura2019MNRAS.485..163K}. The neutrino production in the super-Eddington accretion phase of the MAD state has been discussed for tidal disruption events \citep{Hayasaki2019ApJ...886..114H}.
%\subsection{Acceleration and cooling of high-energy protons}
%Our understanding of the physics of particle acceleration in relativistic reconnection has greatly advanced in recent years thanks to first-principles particle-in-cell (PIC) simulations (Werner \& Uzdensky 2017; Zhang et al. 2021; Zhang et al. 2023; Chernoglazov et al. 2023). 

In the MAD scenario, the proton number density in the accretion flow is $n_{ \textit{p},\rm{MAD}} = \dot{M}/4\pi m_p RH V_{\rm R,MAD}$, where $V_{\rm R,MAD} = \epsilon V_{\rm ff}$ is the radial velocity of the accretion flow, $V_{\rm ff} = \sqrt{2GM_{\rm BH}/R}$ is free-fall velocity and $\epsilon \lesssim 0.01$ \citep{NIA2003PASJ...55L..69N}. Taking disk height $H = R/2$, we have 
\begin{equation}
\begin{aligned}
       &n_{p,\rm{MAD}} \sim 7.3 \times 10^{12}~\rm{cm}^{-3}\\
       &\left(\frac{\dot{m}}{10}\right)\left(\frac{0.01}{\epsilon} \right)\left(\frac{{M}_{\rm BH}}{3\times 10^{8}M_{\odot}}\right)^{1/2}\left(\frac{R}{30R_g}\right)^{-3/2}.
\end{aligned}
\end{equation}
Here we consider a dissipation site with a radius   $R\sim 30 -60 R_g$ for the MAD scenario. For smaller $R$, the cooling of protons due to synchrotron emission  would be too strong and the neutrino emission will be suppressed (as discussed later). 

The magnetic field of MAD can be obtained by equating the magnetic energy density $B_{\rm MAD}^2/(8\pi)$ with the gravitational force per unit area of the radially accreting mass $GM_{\rm BH}m_p n_{\textit{p},\text{MAD}}H/R^2$ \citep{Hayasaki2019ApJ...886..114H}, which is given by
\begin{equation}
\begin{aligned}
    &B_{\rm MAD} = \sqrt{\frac{2GM\dot{M}}{\epsilon V_{\rm ff}R^3}} 
    \sim 6\times 10^4 \,{\rm G}\\
     & \left(\frac {\dot{m}}{10} \right)^{1/2}
    \left(\frac{\epsilon} {0.01}\right)^{-1/2}\left(\frac{M_{\rm BH}}{3\times 10^8 M_{\odot}} \right)^{3/4} \left( \frac{R}{30R_g} \right)^{-5/4}.
\end{aligned}
\end{equation}
Then the acceleration timescale is given by
\begin{equation}
\begin{aligned}
    &t_{\rm acc,MAD} \approx \frac{\eta r_{\rm L}}{c}
    \simeq 55 \,{\rm s}\left(\frac {\dot{m}}{10} \right)^{-1/2}
    \left(\frac{\epsilon} {0.01}\right)^{1/2}\\
    &\left(\frac{E_p}{100\rm{PeV}}\right)\left(\frac{\eta}{300}\right)\left(\frac{M_{\rm BH}}{3\times 10^8 M_{\odot}} \right)^{-3/4}  \left( \frac{R}{30R_g} \right)^{5/4}.
    \label{taccMAD}
\end{aligned}
\end{equation}
The timescale of diffusion is $t_{\rm diff} \approx R^2/D_{\rm R} \sim 6.4\times 10^{9}(R/30R_g)^2(E_p/2\,{\rm PeV})^{-1}\,$s and the timescale of advection is $t_{\rm fall} \approx R/V_{\rm R} \simeq 1.7\times10^7 (R/30R_g)^{3/2}(\epsilon/0.01)^{-1}(M_{\rm BH}/3\times 10^8M_{\odot})^{-1/2}\,$s respectively. The $pp$ cooling timescale is
\begin{equation}
\begin{aligned}
    t_{pp,\rm{MAD}} &\approx 1/n_{p}{\sigma}_{pp}\kappa_{pp}c\\ &\sim 230s \left(\frac{\dot{m}}{10}\right)^{-1}\left(\frac{M}{3\times10^8 M_{\odot}}\right)^{1/2}\left(\frac{R}{30R_g}\right)^{-3/2},
    \label{tppMAD}
\end{aligned}
\end{equation}
and the proton synchrotron timescale is
\begin{equation}
\begin{aligned}
%    &t_{p,\rm{syn}} = \frac{6 \pi m_p c}{\gamma_p \sigma_T B^2} \left( \frac{m_p}{m_e} \right)^2\\
%    &\approx 173 s\left(\frac{\dot{m}}{40}\right)^{-1}\left(\frac{E_p}{100\rm{PeV}}\right)\left(\frac{M_{BH}}{3\times10^8M_{\odot}}\right)^{-3/2}\left(\frac{R}{60R_g}\right)^{-5/2}.
     &t_{p,\rm{syn,MAD}} = \frac{6 \pi m_p c}{\gamma_p \sigma_T B^2} \left(\frac{m_p}{m_e} \right)^2 \approx 12\,{\rm s}\left(\frac {\dot{m}}{10} \right)^{-1}
    \left(\frac{\epsilon} {0.01}\right)\\
     &\left(\frac{E_p}{100\rm{PeV}}\right)^{-1}\left(\frac{M_{\rm BH}}{3\times 10^8 M_{\odot}} \right)^{-3/2} \left( \frac{R}{30R_g} \right)^{5/2}.
    \label{tsynMAD}
\end{aligned}
\end{equation}

{In the $p\gamma$ process, the energy of high-energy protons and the energy of target photons is related by $E_p \varepsilon_{\gamma} \sim 0.15~\rm{GeV}^2$. Therefore, the energy of target photons is $\varepsilon_{\gamma}\sim 1.5~\rm{eV}$ for protons with $E_p\sim 100~\rm{PeV}$.
The number density of target photons in accretion disk is $n_{\gamma} = L_{\rm disk}/4\pi R^2c\varepsilon_0 \approx 3\times 10^{13}~\rm{cm}^{-3}(\textit{L}_{\rm disk}/4.7 \times 10^{43}~erg~s^{-1})(\varepsilon_{\gamma}/1.5~eV)^{-1} (\textit{R}/30\textit{R}_g)^{-2} $, where $L_{\rm disk}$ is the luminosity of the inner disk at 1.5 eV, which is obtained by Eq.\eqref{mbb}}

Therefore, the timescale of photomeson process ($p\gamma$) can be estimated as 
\begin{equation}
\begin{aligned}
    t_{p\gamma,\rm{MAD}}\approx 1/n_{\gamma}{\sigma}_{p\gamma}\kappa_{p\gamma}c \simeq 10^4\, \rm{s}\left(\frac{\textit{n}_{\gamma}}{3\times 10^{13}~\rm{cm}^{-3}}\right)^{-1},
    \label{tpgammaMAD}
\end{aligned}    
\end{equation}
and the Bethe-Heitler timescale $t_{\rm B-H}$  is
\begin{equation}
\begin{aligned}
    t_{\rm B-H,\rm{MAD}}\approx 1/n_{\gamma}\hat{\sigma}_{\rm BH}c \simeq 1.4\times10^6\,\rm{s}\left(\frac{\textit{n}_{\gamma}}{3\times 10^{13}~\rm{cm}^{-3}}\right)^{-1}.
    \label{tBHMAD}
\end{aligned}    
\end{equation} 
%In such scenario, the plasma beta is fixed at 0.33, and we consider a relatively large emission region for $\mathcal{R} \sim 60$. 

%where $\beta$ is plasma beta, $C_s = \sqrt{k_BT_p/m_p}$ is sound speed. We expect that the thermal protons are at the virial temperature $T_p = \sqrt{GM_{\rm BH}m_p/3R k_{\rm B}} $ 
%Taking the proton number density from the optical depth and sound speed argument presented above, we can estimate the magnetization of the system as
%\begin{equation}
%    \sigma_p = \frac{B^2}{4\pi m_p n_p c^2} \sim 1.3 \left(\frac{R}{30R_g}\right)\left(\frac{1}{\beta}\right),
%\end{equation}
%Protons acceleration via magnetic reconnection is efficient for $\sigma_p \gtrsim 1$. 
%In order to make our assumption on magnetic reconnection of protons acceleration consistent, we use $\beta \sim 1$ as a reference value. The acceleration time scale is given by

Comparing the estimated timescales in Eq.\eqref{tpgammaMAD}, Eq.\eqref{tBHMAD}, Eq.\eqref{tppMAD} and Eq.\eqref{tsynMAD}, we find that the proton synchrotron process dominates the cooling in the MAD scenario for typical parameter values.  By equating the acceleration timescale in Eq.\eqref{taccMAD} and the proton synchrotron timescale in Eq.\eqref{tsynMAD}, we can derive the maximum proton energy,
\begin{equation}
\begin{aligned}
    &E_{p,\rm{max,MAD}} \approx 50\,\rm PeV\,\left(\frac{\dot{\textit m}}{10}\right)^{-1/4}\left(\frac{\textit{R}}{30 \textit{r}_g}\right)^{-5/8}\\
    &\left(\frac{M_{\rm BH}}{3\times 10^8 M_{\odot}}\right)^{3/8}
    \left(\frac{\eta}{300}\right)^{-1/2}.
    \label{EpmaxMAD}
\end{aligned}
\end{equation} 

We show the various timescales in the panel (a) of Fig.~\ref{fig:MAD Flare} for $R=30R_g$ and $R=60R_g$ with super-Eddington accretion $\dot{m}=10$. We used strict expressions of $p\gamma$ and Bethe-Heitler process 
\begin{equation}
    t^{-1}_{p\gamma,{\rm B-H}} = \frac{c}{2 \gamma_p^2} \int_{\epsilon_{\text{th}}}^{\infty} \sigma(\bar{\epsilon}) \kappa(\bar{\epsilon}) \bar{\epsilon} \, d\bar{\epsilon} \int_{\bar{\epsilon}/2\gamma_p}^{\infty} \epsilon^{-2} \frac{dn}{d\epsilon}d\epsilon ,
    \label{pgamma timescale}
\end{equation}
in calculating the relevant timescales, $\gamma_p$ is Lorentz factors of protons, $\bar{\epsilon}$ is the photon energy in the proton rest frame, ${dn}/{d\epsilon}$ is the number density of target photons, $\epsilon_{\text{th}}$ is the threshold energy of $p\gamma$ or Bethe-Heitler process,  $\sigma$ is cross-section and $\kappa$ is inelasticity for $p\gamma$ or Bethe-Heitler process.  {  Note that the acceleration timescale required for protons to reach the maximum energy of 50 PeV is on the order of $10^3-10^4{\rm s}$, which is much shorter than the duration of the 2014-2015 flare. Thus, the flare duration does not affect the maximum energy of accelerated protons. } 

For protons energy below 100 TeV, $pp$ collision dominates the cooling process and $p\gamma$ is partly suppressed by the proton synchrotron emission in the energy range of 320 TeV to 32 PeV. From the above timescales, we can derive $pp$ and $p\gamma$ interaction efficiencies, which are shown in the panel (b) of Fig.~\ref{fig:MAD Flare}.
The efficiencies of $pp$ and $p\gamma$ process in the relevant proton energy range are  roughly $f_{pp}\sim 0.9 $ and $f_{p\gamma}\sim 0.01 $ for typical parameter values, respectively. 
Hence, during the 2014-2015 neutrino flare period, the neutrino production is predominantly produced by $pp$ collisions.  

%For the MAD scenario, due to a high magnetic field, the effect of pion cooling is important. 
In panel (c) of Fig.~\ref{fig:MAD Flare}, we show the neutrino spectrum in comparison with the observations by IceCube during the 2014-2015 neutrino flare period of TXS 0506+056.
The solid line represents the neutrino spectrum for $R = 60R_g$,  whereas the dashed line represents the neutrino spectrum for $R = 30R_g$. A smaller radius of the dissipation site leads to a larger magnetic field, thereby increasing the cooling of pions as well as protons and leading to a lower  cutoff energy in the neutrino spectrum. Note that here it is the  pion cooling that determines the cutoff energy in the neutrino spectrum, since the pion cooling is stronger than the proton cooling at high energies for typical parameter values. 

We also consider the steady-state neutrino emission of TXS 0506+056 in the MAD scenario, where a lower accretion rate is applicable. We consider a sub-Eddington accretion rate with $\dot{m}\sim 0.1$ for TXS 0506+056. The corresponding timescales, the $pp$ and $p\gamma$ efficiency, and the neutrino spectrum are shown in  Fig.~\ref{fig:MAD Steady}. We find that $pp$ process still dominates the $p\gamma$ process for typical parameter values and this scenario can explain the 10-year time-integrated neutrino emission observed by IceCube \citep{2022Sci...378..538I}. 
\begin{figure}

\subfigure[]{
\includegraphics[width=0.45\textwidth]{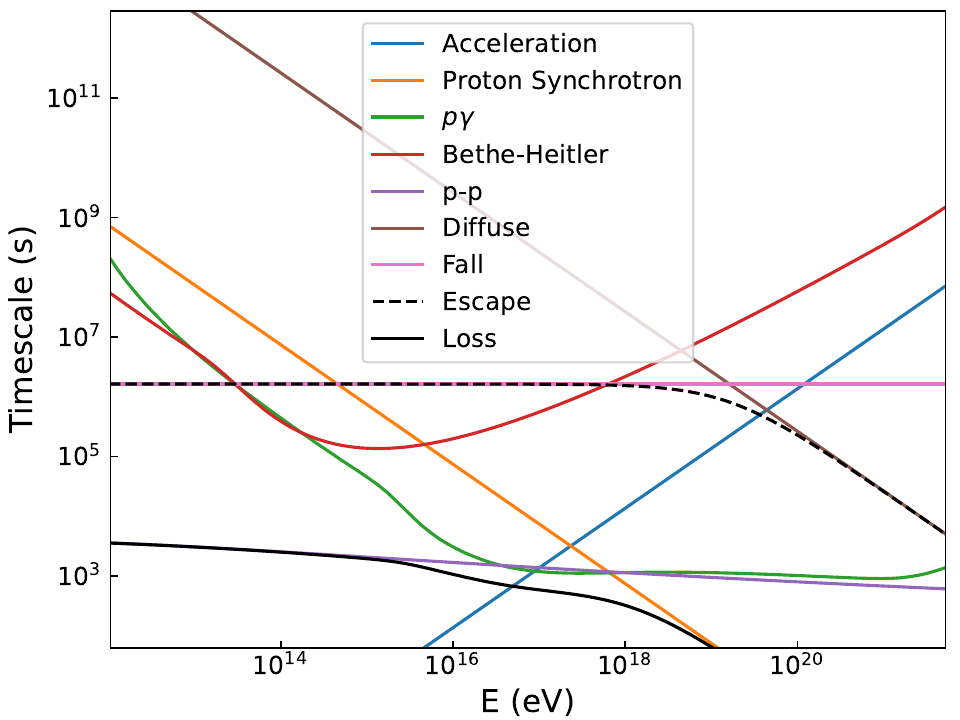}}
\subfigure[]{
\includegraphics[width=0.45\textwidth]{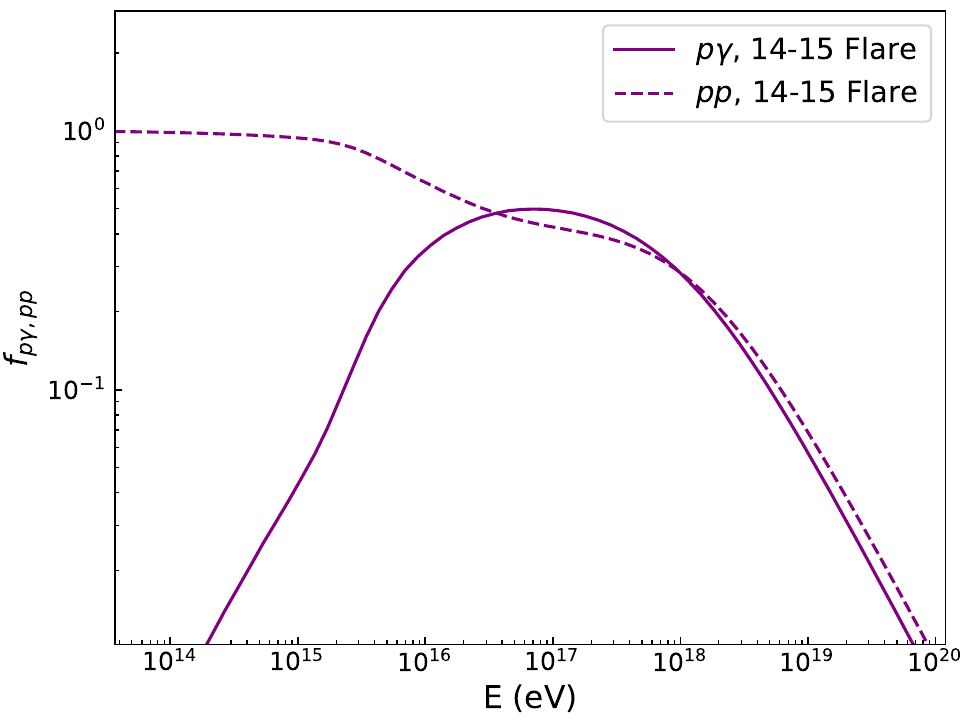}}
\subfigure[]{
\includegraphics[width=0.45\textwidth]{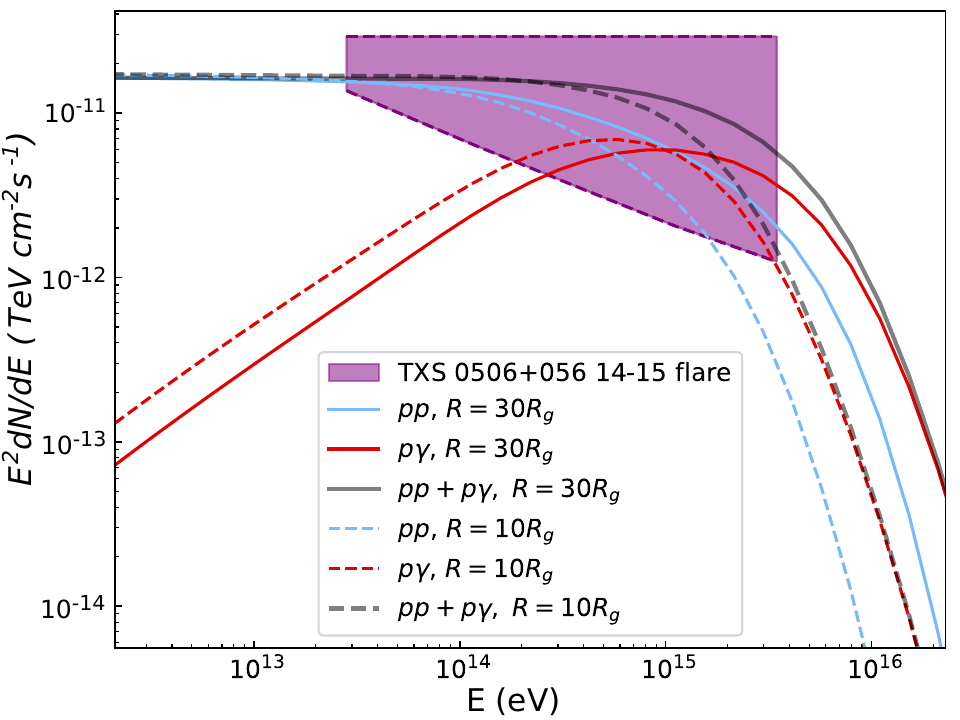}}
\caption{The panels, from top to  bottom, show the various timescales, the efficiencies of $pp$ and $p\gamma$ interactions, and the neutrino spectrum in the SANE scenario. We adopt the parameters $M_{\rm BH} = 3\times10^8 M_{\odot}$, $\epsilon_{\rm CR} = 0.1$,  $\alpha = 0.3$, $\beta = 10$, $\eta=300$, $\beta=10$ and $\dot{m} =10$. For panel (a) and panel (b), the dissipation radius is set as $30R_g$, while in panel (c), two radius with $R =10R_g$ and $R =30 R_g$ are assumed. In panel (c), the observed neutrino spectrum of TXS 0506+056 during the 2014-2015 neutrino flare is also shown.}
\label{fig:SANE Flare}
\end{figure}

\begin{figure}
\centering
\subfigure[]{
\includegraphics[width=0.45\textwidth]{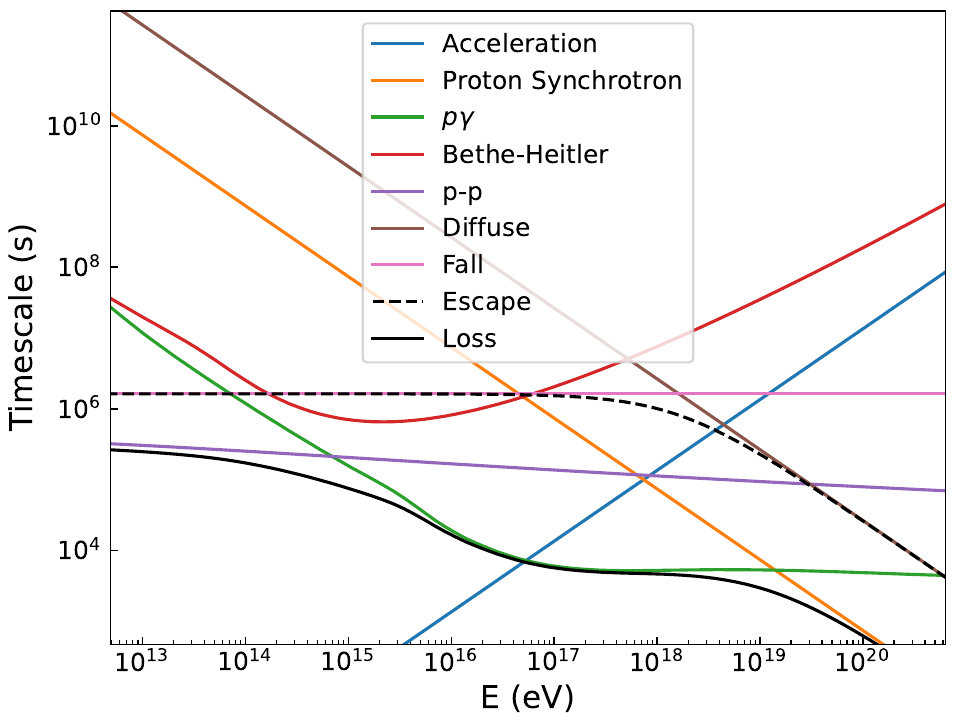}}
\subfigure[]{
\includegraphics[width=0.45\textwidth]{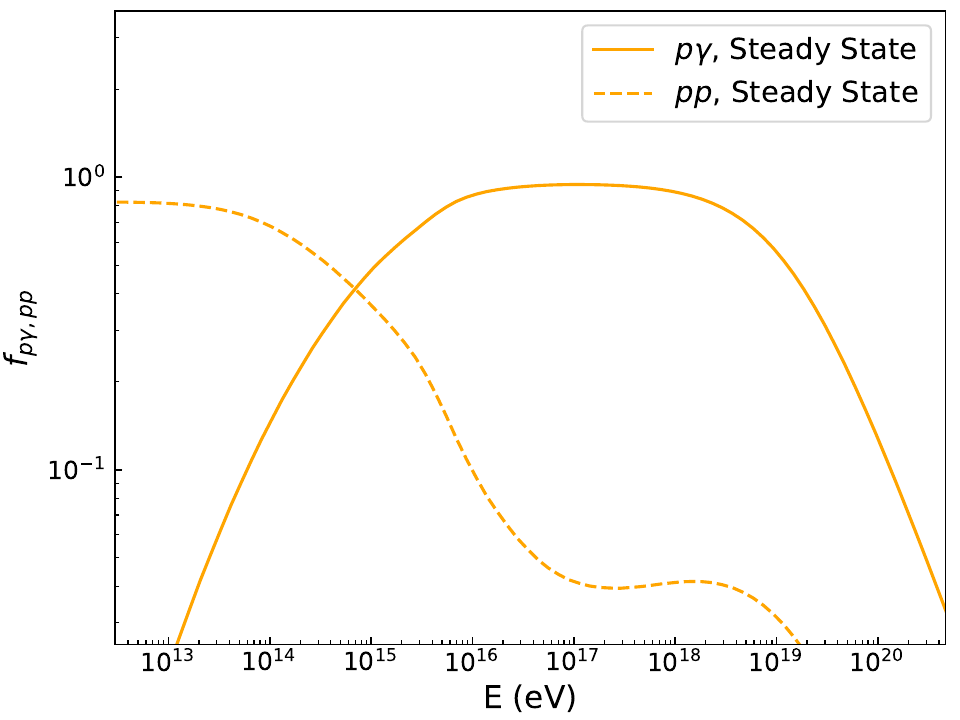}}
\subfigure[]{
\includegraphics[width=0.45\textwidth]{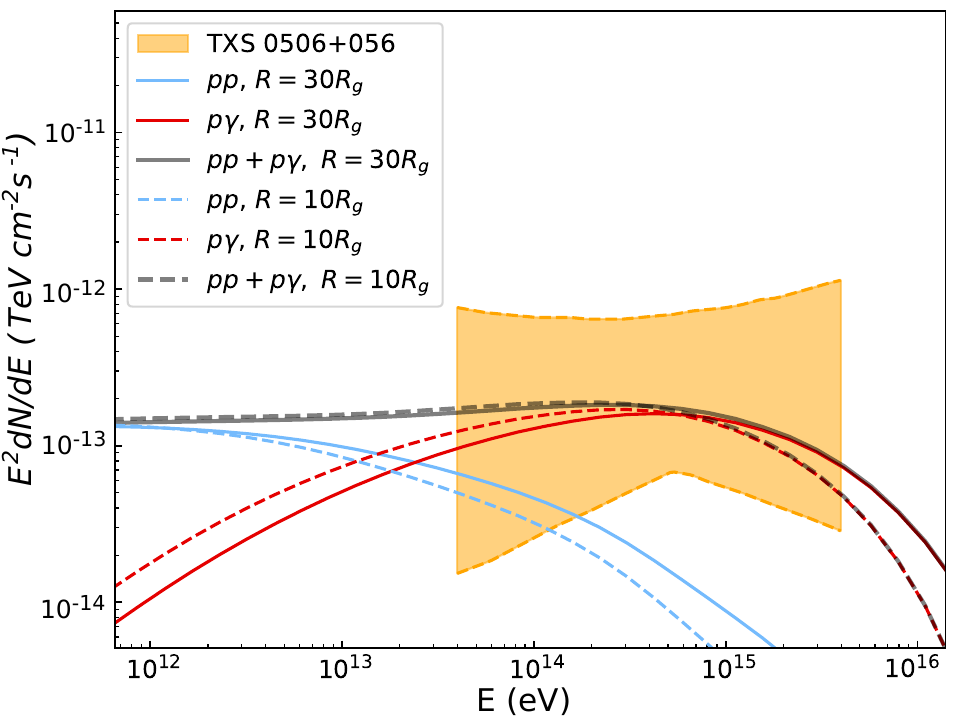}}
\caption{Same as Fig.~\ref{fig:SANE Flare}, but assuming $\dot{m} = 0.1$ to explain the time-integrated neutrino emission of TXS 0506+056 \citep{2022Sci...378..538I}.}
\label{fig:SANE Steady}
\end{figure}

\subsection{The SANE scenario}
In the SANE scenario, a lower magnetic field is expected in the accretion flow.  We set the plasma $\beta$ as $\beta = 10$. The proton number density in the accretion flow is
\begin{equation}
\begin{aligned}
       & n_{p,\rm{SANE}}\sim 2.7 \times 10^{11} \rm{cm}^{-3}\\
       &\left(\frac{\dot{m}}{10}\right)\left(\frac{0.3}{\alpha} \right)\left(\frac{{M}_{\rm BH}}{3\times 10^{8}M_{\odot}}\right)^{1/2}\left(\frac{R}{30R_g}\right)^{-3/2},
       \label{npSANE}
\end{aligned}
\end{equation}
where the radial velocity is $V_{\rm R,SANE} \approx \alpha V_k/2$  in the SANE scenario, $V_k = \sqrt{GM/R}$ is Keplerian velocity and $\alpha \sim 0.3$ is the viscous parameter.  Then the magnetic field is given by
\begin{equation}
\begin{aligned}
    &B_{\rm SANE} = \sqrt{\frac{8\pi n_{p, {\rm SANE}} m_p C_s^2}{\beta}}\sim 2.5\times 10^3 \rm{G}\left(\frac{\dot{\textit m}}{10}\right)^{1/2} \\
    &\left(\frac{\beta}{10}\right)^{-1/2} \left(\frac{M_{\rm BH}}{3\times 10^8 M_{\odot}}\right)^{3/4}\left(\frac{R}{30R_g}\right)^{-5/4}\left(\frac{\alpha}{0.3}\right)^{-1/2},
\end{aligned}
\end{equation}
where $C_s \approx V_k/2$ is sound speed.
The particle acceleration timescale is given by
\begin{equation}
\begin{aligned}
    &t_{\rm acc,SANE} \simeq 1.3\times10^3 \,{\rm s}\left(\frac{\dot{m}}{10}\right)^{-1/2}\left(\frac{\beta}{10}\right)^{1/2}\\ &\left(\frac{E_p}{100\rm{PeV}}\right)\left(\frac{\eta}{300}\right)
    \left(\frac{M_{\rm BH}}{3\times 10^8 M_{\odot}}\right)^{-3/4}\left(\frac{R}{30R_g}\right)^{5/4}\left(\frac{\alpha}{0.3}\right)^{1/2}.
    \label{taccSANE}
\end{aligned}
\end{equation}
The timescale of diffusion  is $t_{\rm diff} \approx R^2/D_{\rm R} \sim 2.6\times 10^{8} {\rm{s} }(R/30R_g)^2(E_p/100~\rm{PeV})^{-1}$  and $t_{\rm fall} \approx R/V_{\rm R} \simeq 1.6\times10^6 {\rm{s}}(R/30R_g)^{3/2}(\alpha/0.3)^{-1}(M_{\rm BH}/3\times 10^8M_{\odot})^{-1/2}$ respectively. 
The  cooling timescales of $pp$ collision and proton synchrotron emission are, respectively,
\begin{equation}
\begin{aligned}
%    t_{pp} &\approx 1/n_{p}\hat{\sigma}_{pp}\kappa_{pp}c\\ &\sim 1.6\times10^3s\left(\frac{\dot{m}}{40}\right)^{-1}\left(\frac{M}{3\times10^8 M_{\odot}}\right)^{1/2}\left(\frac{R}{60R_g}\right)^{-3/2},
    &t_{pp,\rm{SANE}} \sim 6\times10^3 \,{\rm s}\\
    &\left(\frac{\dot{m}}{10}\right)^{-1}\left(\frac{\alpha}{0.3}\right)\left(\frac{{M}_{\rm BH}}{3\times 10^{8}M_{\odot}}\right)^{-1/2}\left(\frac{R}{30R_g}\right)^{3/2}
    \label{tppSANE}
\end{aligned}
\end{equation}
and
\begin{equation}
\begin{aligned}
%    &t_{p,\rm{syn}} = \frac{6 \pi m_p c}{\gamma_p \sigma_T B^2} \left( \frac{m_p}{m_e} \right)^2\\
%    &\approx 173 s\left(\frac{\dot{m}}{40}\right)^{-1}\left(\frac{E_p}{100\rm{PeV}}\right)\left(\frac{M_{BH}}{3\times10^8M_{\odot}}\right)^{-3/2}\left(\frac{R}{60R_g}\right)^{-5/2}.
     &t_{p,\rm{syn,SANE}} \sim 7\times 10^3\,{\rm s} \left(\frac{\dot{m}}{10}\right)^{-1}\left(\frac{\beta}{10}\right) \\
     &\left(\frac{E_p}{100\rm{PeV}}\right)^{-1}
    \left(\frac{M_{\rm BH}}{3\times 10^8 M_{\odot}}\right)^{-3/2}\left(\frac{R}{30R_g}\right)^{5/2}\left(\frac{\alpha}{0.3}\right).
    \label{tsynSANE}
\end{aligned}
\end{equation}
For the photon field in the SANE scenario, the number density of target photons  is estimated to be $n_{\gamma} = L_{\rm disk}/4\pi R^2c\varepsilon_0 \approx 3\times 10^{13}~\rm{cm}^{-3}(\textit{L}_{\rm disk}/4.7 \times 10^{43}~erg~s^{-1})(\varepsilon_0/1.5~eV)^{-1} (\textit{R}/30\textit{R}_g)^{-2} $.
Then the timescale of photomeson process ($p\gamma$) can be estimated as 
\begin{equation}
%\begin{aligned}
    t_{p\gamma,\rm{SANE}} \simeq 10^4 \,{\rm s}\left(\frac{n_{\gamma}}{3\times 10^{13}~\rm{cm}^{-3}}\right)^{-1},
    \label{tpgammaSANE}
%\end{aligned}    
\end{equation}
and the timescale of Bethe-Heitler is 
\begin{equation}
%\begin{aligned}
    t_{\rm B-H,SANE} \simeq 1.4\times10^6\,{\rm s}\left(\frac{n_{\gamma}}{3\times 10^{13}~\rm{cm}^{-3}}\right)^{-1}.
    \label{tBHSANE}
%\end{aligned}    
\end{equation}

From the timescales above, we find that  in the SANE scenario, depending on the parameter values, the $ pp $, $p\gamma$ and the proton synchrotron emission processes all could become the dominant cooling mechanism for the highest energy protons. If the $pp$ process becomes the dominant mechanism for the proton cooling, by equating the acceleration timescale with the $pp$ cooling timescale, we  obtain the maximum proton energy, 
\begin{equation}
\begin{aligned}
    &E_{p,\rm{max,SANE}} \approx 400\,{\rm PeV}\\ &\left(\frac{\dot{m}}{10}\right)^{-1/2}\left(\frac{M_{\rm BH}}{3\times 10^8 M_{\odot}}\right)^{1/4}\left(\frac{R}{30R_g}\right)^{1/4}\left(\frac{\beta}{10}\right)^{-1/2}\\
    &\left(\frac{\alpha}{0.3}\right)^{1/2}\left(\frac{\eta}{300}\right)^{-1}. 
    \label{EpmaxSANE1}
\end{aligned}
\end{equation}
Similarly, if the $p\gamma$ process becomes the predominant mechanism for the proton cooling, the maximum energy is
\begin{equation}
\begin{aligned}
    &E_{p,\rm{max,SANE}} \approx 800\,{\rm PeV}\\ &\left(\frac{\dot{m}}{10}\right)^{1/2}\left(\frac{M_{\rm BH}}{3\times 10^8 M_{\odot}}\right)^{3/4}\left(\frac{R}{30R_g}\right)^{-5/4}\left(\frac{\beta}{10}\right)^{-1/2}\\
    &\left(\frac{\alpha}{0.3}\right)^{-1/2}\left(\frac{n_{\gamma}}{3\times 10^{13}~\rm{cm}^{-3}}\right)^{-1}\left(\frac{\eta}{300}\right)^{-1},
    \label{EpmaxSANE2}
\end{aligned}
\end{equation}
and if the proton synchrotron emission becomes the predominant mechanism for the proton cooling, the maximum energy is
\begin{equation}
\begin{aligned}
    &E_{p,\rm{max,SANE}} = 250\,{\rm PeV}\\ 
    &\left(\frac{\dot{m}}{10}\right)^{-1/4}\left(\frac{\beta}{10}\right)^{1/4} 
    \left(\frac{M_{\rm BH}}{3\times 10^8 M_{\odot}}\right)^{-3/8}\left(\frac{R}{30R_g}\right)^{5/8}\\
    &\left(\frac{\alpha}{0.3}\right)^{1/4}\left(\frac{\eta}{300}\right)^{-1/2}.
    \label{EpmaxSANE3}
\end{aligned}
\end{equation}

The panel (a) of Fig.~\ref{fig:SANE Flare} shows various timescales in the SANE scenario for a dissipation radius of $R = 30R_g$.  For low-energy protons, the cooling is dominantly by $pp$ interactions, whereas for the high-energy protons the cooling of both $p\gamma$ interaction and proton synchrotron emission are important. The efficiency for $pp$ and $p\gamma$ are shown in panel (b) of Fig.~\ref{fig:SANE Flare}, which gives $f_{pp} \sim 1$ and $f_{p\gamma} \sim 0.6$ in the  neutrino energy range where the respective cooling process is dominated. In panel (c) of Fig.~\ref{fig:SANE Flare}, we show the neutrino spectrum in comparison with the observation data during the 2014-2015 neutrino flare period of TXS 0506+056.   Owing to a reduced magnetic field in the SANE scenario, the size of the dissipation region could be considerably smaller without suffering from a strong cooling for protons.  Therefore we  consider two dissipation radii with $R =10R_g$ and $R =30R_g$, respectively. We find that $pp$ and $p\gamma$   processes contribute significantly to the neutrino flux at lower and higher energies, respectively. 

Same as the MAD scenario, we also consider the steady-state neutrino emission of TXS 0506+056  assuming a low accretion rate $\dot{m}\sim 0.1$. The corresponding timescales, the $pp$ and $p\gamma$ efficiencies, and the neutrino spectrum are shown in  Fig.~\ref{fig:SANE Steady}. Due to a lower density in the accretion flow, the $pp$ efficiency becomes lower. As a result, the $p\gamma$   process becomes dominant in the neutrino production in the observed energy range by IceCube. 
%Insofar we take $\eta = 30000$, for both MAD and SANE scenario
\begin{figure*}
\centering
\subfigure[]{
\includegraphics[width=0.49\textwidth]{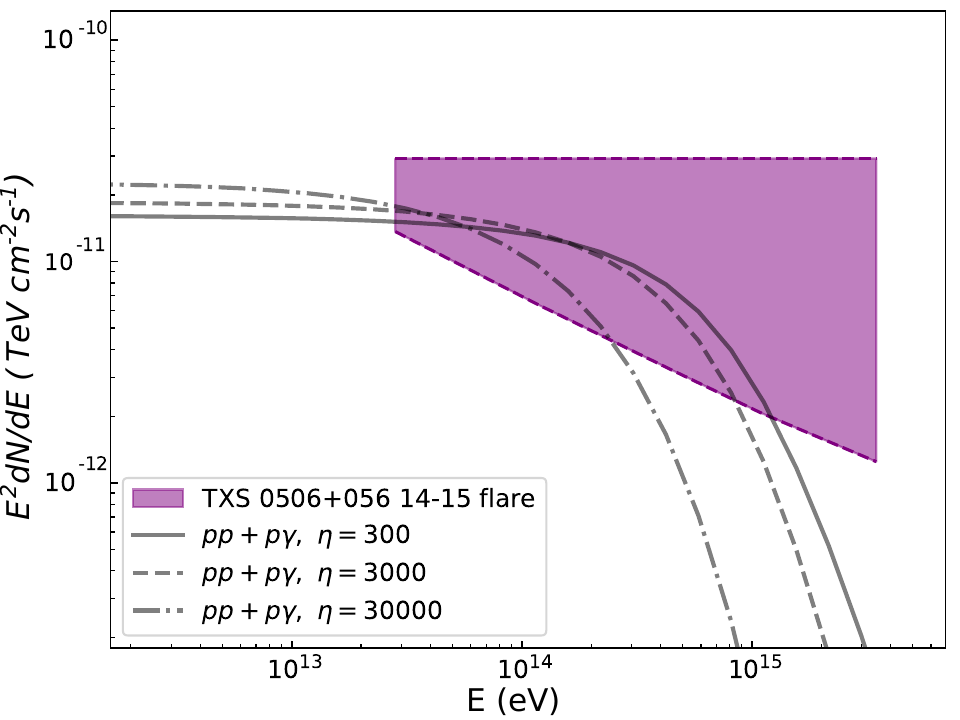}}
\subfigure[]{
\includegraphics[width=0.49\textwidth]{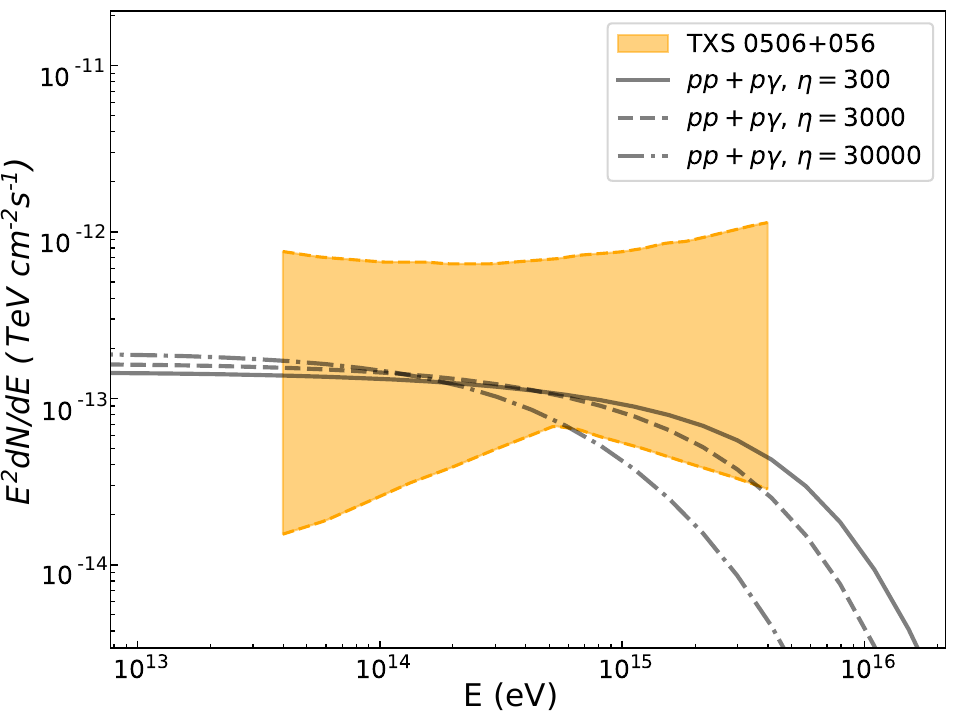}}
\caption{The neutrino spectrum with different acceleration efficiency $\eta$ in the MAD scenario. The parameter values used are:  $R = 60R_g$, $M_{\rm BH} = 3\times10^8 M_{\odot}$, $\epsilon_{\rm CR} = 0.1$,  and $\epsilon = 0.01$. The solid line, the dashed line and the dot-dashed line represent $\eta = 300$, $\eta = 3000$ and $\eta = 30000$, respectively. The left panel shows the super-Eddington regime ($\dot{m}= 10$) for the 2014-2015 neutrino flare of TXS 0506+056, whereas the right panel shows the sub-Eddington accretion regime ($\dot{m}= 0.1$) for the steady-state emission. }
\label{fig:MAD eta}
\end{figure*}

\begin{figure*}
\centering
\subfigure[]{
\includegraphics[width=0.49\textwidth]{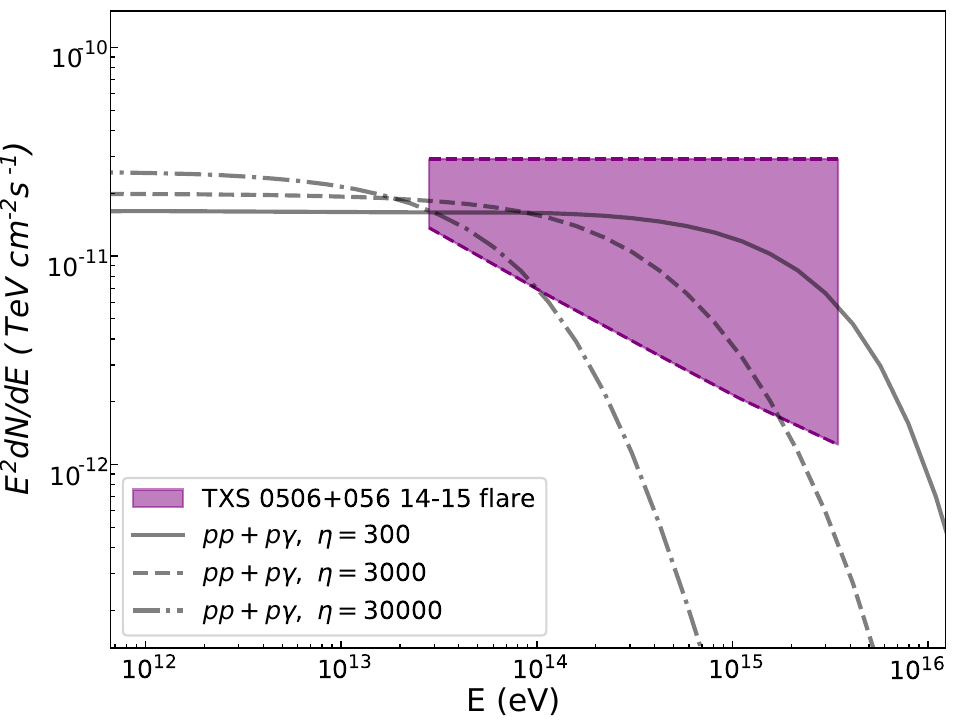}}
\subfigure[]{
\includegraphics[width=0.49\textwidth]{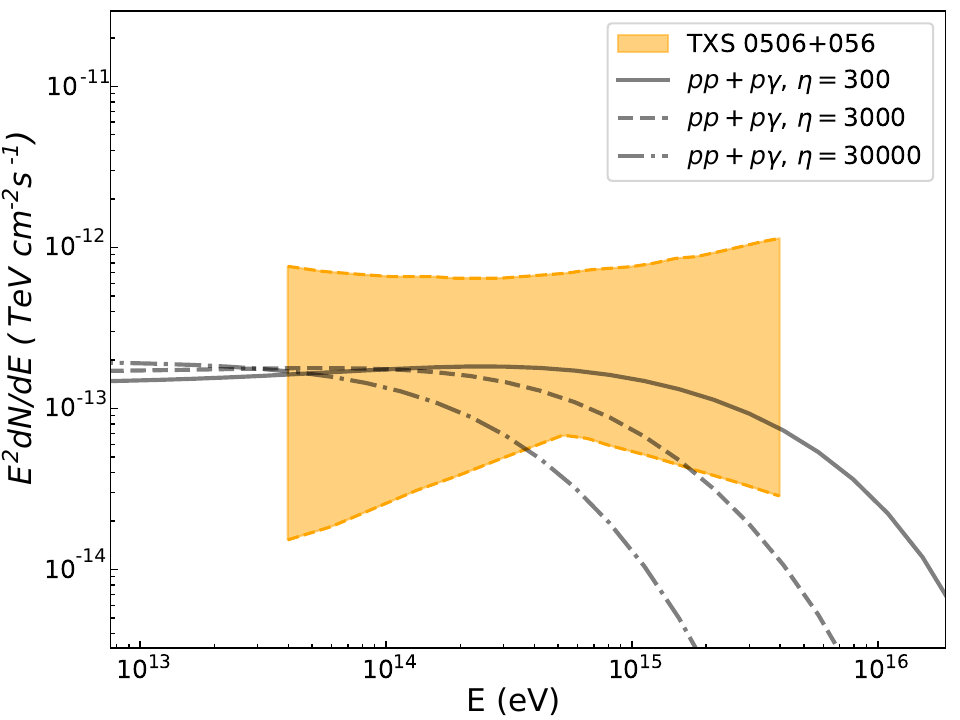}}
\caption{Same as Fig.~\ref{fig:MAD eta}, but for the SANE scenario. The parameter values used are: $R = 30R_g$, $M_{\rm BH} = 3\times10^8 M_{\odot}$, $\epsilon_{\rm CR} = 0.1$,  $\alpha = 0.3$, and $\beta = 10$. }
\label{fig:SANE eta}
\end{figure*}

\section{Discussions}
We have assumed $\eta=300$ in the above calculations. However, the value of $\eta$ depends on the specific acceleration mechanism. If the acceleration efficiency is lower, $\eta$ could be significantly larger. We study its influence on the neutrino spectrum in the MAD and SANE scenarios by replacing $\eta=300$ with $\eta=3000$ and $\eta=30000$ in the calculation. 
Fig.~\ref{fig:MAD eta} and Fig.~\ref{fig:SANE eta} show the neutrino spectrum in the MAD and SANE scenarios with various acceleration efficiency $\eta$, respectively. In the super-Eddington accretion MAD case (the left panel of Fig.~\ref{fig:MAD eta}), we find that the neutrino spectrum remains almost unchanged between  $\eta=300$ and $\eta=3000$. This is because the pion cooling is still dominated over the proton synchrotron cooling, although $E_{p,\rm{max}}$ decreases by a factor of 3. However, for $\eta=30000$,  the neutrino spectrum softens significantly following the decrease of $E_{p,\rm{max}}$.  In the  sub-Eddington accretion MAD case (the right panel of Fig.~\ref{fig:MAD eta}), the situation is quite similar, although the  cutoff energy of the neutrino spectrum is higher  due to a larger $E_{p,\rm{max}}$ for smaller accretion rate (see Eq.\eqref{EpmaxMAD}). 
In the SANE scenario, as $\eta$ increases and $E_{p,\rm{max}}$ decreases,   the  cutoff energy of the neutrino spectrum  decreases almost in a linear relation with $\eta$ in both super-Eddington accretion and sub-Eddington accretion regimes. For $\eta=30000$, the neutrino spectrum is too soft to explain the data. Therefore, to explain the observed neutrino spectrum of TXS 0506+056,  we need an acceleration efficiency  $\eta <30000$ in both MAD and SANE scenarios.
%Setting $\eta = 30000$ results in reduced acceleration efficiency, subsequently decreasing the maximum energy $E_{p,\rm{max}}$. This poses a challenge in explaining the observed neutrino flux from TXS 0506+056.

{  GeV-TeV gamma-rays are observed from TXS 0506+056. Based on the $\gamma\gamma$ optical depth calculations by \citep{Murase2020PhRvL.125a1101M}, $\gamma\gamma$ absorption in the compact corona and accretion disk can effectively suppress GeV-TeV gamma-ray emission from hadronic processes (i.e. the so-called hidden gamma-ray source). The observed GeV-TeV gamma rays from TXS 0506+056 may then arise from the jets, as usually assumed for GeV-TeV blazars \citep{Madejski2016ARA&A..54..725M, Gao2019NatAs...3...88G, Liu2023MNRAS.526.5054L}. These emissions are produced through synchrotron self-Compton (SSC)  and/or external Compton (EC)  processes of relativistic leptons. Being far from the compact corona/disk, these high-energy gamma-rays can avoid absorption and become detectable. 

Most previous studies have suggested that the high-energy neutrinos from TXS 0506+056 originate from the relativistic jet in a region with a radius much larger than that of the accretion flow \citep{Gao2019NatAs...3...88G,Keivani2018,Cerruti2019}. One way to distinguish between our accretion flow model and the jet model is through the cascade emission induced by the absorbed high-energy 
gamma-rays accompanying the neutrino production. The compact size of the neutrino production region in the accretion flow model leads to strong absorption of the high-energy gamma-rays and the absorbed energy is transferred to MeV emission through the electromagnetic cascades\citep{Murase2020PhRvL.125a1101M}, Future mission such as AMEGO-X \citep{Caputo2022JATIS...8d4003C} and eASTROGAM \citep{De2017ExA....44...25D} can 
detect the MeV emission and test the accretion flow model. In addition, our model predicts that mis-aligned blazars (i.e., radio galaxies), where the jets do not point toward us,  can also produce high-energy neutrinos through the accretion flow. Future observations by next-generation neutrino telescopes, such as IceCube-Gen2 \citep{Lu2023APS..APRD13008L} and HUNT \citep{2024icrc.confE1080H} can test this.
}

\section{Summary}
In this work, we study whether the neutrino emission from TXS 0506+056   could come from the accretion flow, instead of the usually discussed relativistic jet. We find that a super-Eddington accretion with 
$\dot{M}\sim 10 \dot{M}_{\rm Edd}$ is needed to explain the neutrino outburst during 2014-2015. The accretion flow may also produce the long-term neutrino emission during the steady state when the accretion drops to sub--Eddington rate. The accretion flow could be a MAD with highly magnetized plasma. Magnetic reconnections and/or plasma turbulence in the MAD may  accelerate  cosmic ray particles, which produce neutrinos via $pp$ and $p\gamma$ processes.   Compared with the SANE accretion flow,  the MAD has a higher magnetic field, which leads to stronger cooling of  cosmic ray protons and secondary pions. As a result, a larger radius for the dissipation in the MAD scenario is needed to avoid this cooling effect.   The size of the neutrino production site is still sufficiently compact so that the TeV-PeV gamma-rays accompanied the neutrinos are absorbed by the dense optical to X-ray photons in the AGN core region.

In a super-Eddington accretion flow, $pp$ interactions play a dominant role in producing neutrinos because of the high density of the accretion flow. This leads to a flat neutrino spectrum with a high-energy cutoff, which is different from the  spectrum of neutrino emission produced in the $p\gamma$ process. This may explain the hard neutrino spectrum in TXS 0506+056 during 2014-2015, in contrast to the soft spectrum of neutrino emission of NGC 1068, which is usually thought to be produced by the $p\gamma$ process.

\section{Acknowledgements}
We would like to thank Haoning He for helpful discussions on the long-term neutrino flux from TXS 0506+056.  This work is
supported by the National Natural Science Foundation of China (grant Nos. 12333006 and 12121003, 12393852). We are grateful to the High Performance Computing Center (HPCC) of Nanjing University for doing the numerical calculations in this paper on its blade cluster system.

Note added.— While we were finalizing this manuscript,
we became aware of the work of \cite{zathul2024ngc1068informedunderstandingneutrino}
 (arXiv:24.14598), which also propose that the neutrino emission from TXS 0506+056 could originate near its AGN core.

\bibliography{main.bib}{}

\begin{thebibliography}{}
\expandafter\ifx\csname natexlab\endcsname\relax\def\natexlab#1{#1}\fi
\providecommand{\url}[1]{\href{#1}{#1}}
\providecommand{\dodoi}[1]{doi:~\href{http://doi.org/#1}{\nolinkurl{#1}}}
\providecommand{\doeprint}[1]{\href{http://ascl.net/#1}{\nolinkurl{http://ascl.net/#1}}}
\providecommand{\doarXiv}[1]{\href{https://arxiv.org/abs/#1}{\nolinkurl{https://arxiv.org/abs/#1}}}

\bibitem[{{Acciari} {et~al.}(2022){Acciari}, {Aniello}, {Ansoldi}, {Antonelli}, {Arbet Engels}, {Artero}, {Asano}, {Baack}, {Babi{\'c}}, {Baquero}, {Barres de Almeida}, {Barrio}, {Batkovi{\'c}}, {Becerra Gonz{\'a}lez}, {Bednarek}, {Bernardini}, {Bernardos}, {Berti}, {Besenrieder}, {Bhattacharyya}, {Bigongiari}, {Biland}, {Blanch}, {B{\"o}kenkamp}, {Bonnoli}, {Bo{\v{s}}njak}, {Busetto}, {Carosi}, {Ceribella}, {Cerruti}, {Chai}, {Chilingarian}, {Cikota}, {Colombo}, {Contreras}, {Cortina}, {Covino}, {D'Amico}, {D'Elia}, {Vela}, {Dazzi}, {De Angelis}, {De Lotto}, {Del Popolo}, {Delfino}, {Delgado}, {Mendez}, {Depaoli}, {Di Pierro}, {Di Venere}, {Do Souto Espi{\~n}eira}, {Dominis Prester}, {Donini}, {Dorner}, {Doro}, {Elsaesser}, {Fallah Ramazani}, {Fari{\~n}a}, {Fattorini}, {Font}, {Fruck}, {Fukami}, {Fukazawa}, {Garc{\'\i}a L{\'o}pez}, {Garczarczyk}, {Gasparyan}, {Gaug}, {Giglietto}, {Giordano}, {Gliwny}, {Godinovi{\'c}}, {Green}, {Green}, {Hadasch}, {Hahn}, {Hassan}, {Heckmann}, {Herrera}, {Hoang}, {Hrupec}, {H{\"u}tten}, {Inada}, {Iotov}, {Ishio}, {Iwamura}, {Jim{\'e}nez Mart{\'\i}nez}, {Jormanainen}, {Jouvin}, {Kerszberg}, {Kobayashi}, {Kubo}, {Kushida}, {Lamastra}, {Lelas}, {Leone}, {Lindfors}, {Linhoff}, {Lombardi}, {Longo}, {L{\'o}pez-Coto}, {L{\'o}pez-Moya}, {L{\'o}pez-Oramas}, {Loporchio}, {Machado de Oliveira Fraga}, {Maggio}, {Majumdar}, {Makariev}, {Mallamaci}, {Maneva}, {Manganaro}, {Mannheim}, {Mariotti}, {Mart{\'\i}nez}, {Mas Aguilar}, {Mazin}, {Menchiari}, {Mender}, {Mi{\'c}anovi{\'c}}, {Miceli}, {Miener}, {Miranda}, {Mirzoyan}, {Molina}, {Moralejo}, {Morcuende}, {Moreno}, {Moretti}, {Nakamori}, {Nava}, {Neustroev}, {Nievas Rosillo}, {Nigro}, {Nilsson}, {Nishijima}, {Noda}, {Nozaki}, {Ohtani}, {Oka}, {Otero-Santos}, {Paiano}, {Palatiello}, {Paneque}, {Paoletti}, {Paredes}, {Pavleti{\'c}}, {Pe{\~n}il}, {Persic}, {Pihet}, {Prada Moroni}, {Prandini}, {Priyadarshi}, {Puljak}, {Rhode}, {Rib{\'o}}, {Rico}, {Righi}, {Rugliancich}, {Sahakyan}, {Saito}, {Sakurai}, {Satalecka}, {Saturni}, {Schleicher}, {Schmidt}, {Schmuckermaier}, {Schweizer}, {Sitarek}, {{\v{S}}nidari{\'c}}, {Sobczynska}, {Spolon}, {Stamerra}, {Stri{\v{s}}kovi{\'c}}, {Strom}, {Strzys}, {Suda}, {Suri{\'c}}, {Takahashi}, {Takeishi}, {Tavecchio}, {Temnikov}, {Terzi{\'c}}, {Teshima}, {Tosti}, {Truzzi}, {Tutone}, {Ubach}, {van Scherpenberg}, {Vanzo}, {Vazquez Acosta}, {Ventura}, {Verguilov}, {Viale}, {Vigorito}, {Vitale}, {Vovk}, {Will}, {Wunderlich}, {Yamamoto}, {Zari{\'c}}, {Hodges}, {Hovatta}, {Kiehlmann}, {Liodakis}, {Max-Moerbeck}, {Pearson}, {Readhead}, {Reeves}, {L{\"a}hteenm{\"a}ki}, {Tornikoski}, {Tammi}, {D'Ammando}, \& {Marchini}}]{Acciari2022ApJ...927..197A}
{Acciari}, V.~A., {Aniello}, T., {Ansoldi}, S., {et~al.} 2022, \apj, 927, 197, \dodoi{10.3847/1538-4357/ac531d}

\bibitem[{{Ball} {et~al.}(2018){Ball}, {{\"O}zel}, {Psaltis}, {Chan}, \& {Sironi}}]{Ball2018ApJ...853..184B}
{Ball}, D., {{\"O}zel}, F., {Psaltis}, D., {Chan}, C.-K., \& {Sironi}, L. 2018, \apj, 853, 184, \dodoi{10.3847/1538-4357/aaa42f}

\bibitem[{{Banik} {et~al.}(2020){Banik}, {Bhadra}, {Pandey}, \& {Majumdar}}]{Banik2020}
{Banik}, P., {Bhadra}, A., {Pandey}, M., \& {Majumdar}, D. 2020, \prd, 101, 063024, \dodoi{10.1103/PhysRevD.101.063024}

\bibitem[{{Bauer} {et~al.}(2015){Bauer}, {Ar{\'e}valo}, {Walton}, {Koss}, {Puccetti}, {Gandhi}, {Stern}, {Alexander}, {Balokovi{\'c}}, {Boggs}, {Brandt}, {Brightman}, {Christensen}, {Comastri}, {Craig}, {Del Moro}, {Hailey}, {Harrison}, {Hickox}, {Luo}, {Markwardt}, {Marinucci}, {Matt}, {Rigby}, {Rivers}, {Saez}, {Treister}, {Urry}, \& {Zhang}}]{2015ApJ...812..116B}
{Bauer}, F.~E., {Ar{\'e}valo}, P., {Walton}, D.~J., {et~al.} 2015, \apj, 812, 116, \dodoi{10.1088/0004-637X/812/2/116}

\bibitem[{{Bisnovatyi-Kogan} \& {Ruzmaikin}(1974)}]{Bisnovatyi1974Ap&SS..28...45B}
{Bisnovatyi-Kogan}, G.~S., \& {Ruzmaikin}, A.~A. 1974, \apss, 28, 45, \dodoi{10.1007/BF00642237}

\bibitem[{{Caputo} {et~al.}(2022){Caputo}, {Ajello}, {Kierans}, {Perkins}, {Racusin}, {Baldini}, {Baring}, {Bissaldi}, {Burns}, {Cannady}, {Charles}, {da Silva}, {Fang}, {Fleischhack}, {Fryer}, {Fukazawa}, {Grove}, {Hartmann}, {Howell}, {Jadhav}, {Karwin}, {Kocevski}, {Kurahashi}, {Latronico}, {Lewis}, {Leys}, {Lien}, {Marcotulli}, {Martinez-Castellanos}, {Mazziotta}, {McEnery}, {Metcalfe}, {Murase}, {Negro}, {Parker}, {Phlips}, {Prescod-Weinstein}, {Razzaque}, {Shawhan}, {Sheng}, {Shutt}, {Shy}, {Sleator}, {Steinhebel}, {Striebig}, {Suda}, {Tak}, {Tajima}, {Valverde}, {Venters}, {Wadiasingh}, {Woolf}, {Wulf}, {Zhang}, \& {Zoglauer}}]{Caputo2022JATIS...8d4003C}
{Caputo}, R., {Ajello}, M., {Kierans}, C.~A., {et~al.} 2022, Journal of Astronomical Telescopes, Instruments, and Systems, 8, 044003, \dodoi{10.1117/1.JATIS.8.4.044003}

\bibitem[{{Celotti} \& {Blandford}(2001)}]{2001bhbg.conf..206C}
{Celotti}, A., \& {Blandford}, R.~D. 2001, in Black Holes in Binaries and Galactic Nuclei, ed. L.~{Kaper}, E.~P.~J. V.~D. {Heuvel}, \& P.~A. {Woudt}, 206, \dodoi{10.1007/10720995_43}

\bibitem[{{Cerruti} {et~al.}(2019){Cerruti}, {Zech}, {Boisson}, {Emery}, {Inoue}, \& {Lenain}}]{Cerruti2019}
{Cerruti}, M., {Zech}, A., {Boisson}, C., {et~al.} 2019, \mnras, 483, L12, \dodoi{10.1093/mnrasl/sly210}

\bibitem[{{De Angelis} {et~al.}(2017){De Angelis}, {Tatischeff}, {Tavani}, {Oberlack}, {Grenier}, {Hanlon}, {Walter}, {Argan}, {von Ballmoos}, {Bulgarelli}, {Donnarumma}, {Hernanz}, {Kuvvetli}, {Pearce}, {Zdziarski}, {Aboudan}, {Ajello}, {Ambrosi}, {Bernard}, {Bernardini}, {Bonvicini}, {Brogna}, {Branchesi}, {Budtz-Jorgensen}, {Bykov}, {Campana}, {Cardillo}, {Coppi}, {De Martino}, {Diehl}, {Doro}, {Fioretti}, {Funk}, {Ghisellini}, {Grove}, {Hamadache}, {Hartmann}, {Hayashida}, {Isern}, {Kanbach}, {Kiener}, {Kn{\"o}dlseder}, {Labanti}, {Laurent}, {Limousin}, {Longo}, {Mannheim}, {Marisaldi}, {Martinez}, {Mazziotta}, {McEnery}, {Mereghetti}, {Minervini}, {Moiseev}, {Morselli}, {Nakazawa}, {Orleanski}, {Paredes}, {Patricelli}, {Peyr{\'e}}, {Piano}, {Pohl}, {Ramarijaona}, {Rando}, {Reichardt}, {Roncadelli}, {Silva}, {Tavecchio}, {Thompson}, {Turolla}, {Ulyanov}, {Vacchi}, {Wu}, \& {Zoglauer}}]{De2017ExA....44...25D}
{De Angelis}, A., {Tatischeff}, V., {Tavani}, M., {et~al.} 2017, Experimental Astronomy, 44, 25, \dodoi{10.1007/s10686-017-9533-6}

\bibitem[{{de Gouveia Dal Pino} {et~al.}(2010){de Gouveia Dal Pino}, {Piovezan}, \& {Kadowaki}}]{Pino2010}
{de Gouveia Dal Pino}, E.~M., {Piovezan}, P.~P., \& {Kadowaki}, L.~H.~S. 2010, \aap, 518, A5, \dodoi{10.1051/0004-6361/200913462}

\bibitem[{{G{\'a}mez Rosas} {et~al.}(2022){G{\'a}mez Rosas}, {Isbell}, {Jaffe}, {Petrov}, {Leftley}, {Hofmann}, {Millour}, {Burtscher}, {Meisenheimer}, {Meilland}, {Waters}, {Lopez}, {Lagarde}, {Weigelt}, {Berio}, {Allouche}, {Robbe-Dubois}, {Cruzal{\`e}bes}, {Bettonvil}, {Henning}, {Augereau}, {Antonelli}, {Beckmann}, {van Boekel}, {Bendjoya}, {Danchi}, {Dominik}, {Drevon}, {Gallimore}, {Graser}, {Heininger}, {Hocd{\'e}}, {Hogerheijde}, {Hron}, {Impellizzeri}, {Klarmann}, {Kokoulina}, {Labadie}, {Lehmitz}, {Matter}, {Paladini}, {Pantin}, {Pott}, {Schertl}, {Soulain}, {Stee}, {Tristram}, {Varga}, {Woillez}, {Wolf}, {Yoffe}, \& {Zins}}]{2022Natur.602..403G}
{G{\'a}mez Rosas}, V., {Isbell}, J.~W., {Jaffe}, W., {et~al.} 2022, \nat, 602, 403, \dodoi{10.1038/s41586-021-04311-7}

\bibitem[{{Gangadhara} \& {Lesch}(1997)}]{Gangadhara97}
{Gangadhara}, R.~T., \& {Lesch}, H. 1997, \aap, 323, L45, \dodoi{10.48550/arXiv.astro-ph/9707182}

\bibitem[{{Gao} {et~al.}(2019){Gao}, {Fedynitch}, {Winter}, \& {Pohl}}]{Gao2019NatAs...3...88G}
{Gao}, S., {Fedynitch}, A., {Winter}, W., \& {Pohl}, M. 2019, Nature Astronomy, 3, 88, \dodoi{10.1038/s41550-018-0610-1}

\bibitem[{{Hayasaki} \& {Yamazaki}(2019)}]{Hayasaki2019ApJ...886..114H}
{Hayasaki}, K., \& {Yamazaki}, R. 2019, \apj, 886, 114, \dodoi{10.3847/1538-4357/ab44ca}

\bibitem[{{Hoshino}(2012)}]{Hoshino2012PhRvL.108m5003H}
{Hoshino}, M. 2012, \prl, 108, 135003, \dodoi{10.1103/PhysRevLett.108.135003}

\bibitem[{{Hoshino}(2013)}]{Hoshino2013}
---. 2013, \apj, 773, 118, \dodoi{10.1088/0004-637X/773/2/118}

\bibitem[{{Huang} {et~al.}(2020){Huang}, {Luo}, {Du}, {Hu}, {Wang}, \& {Li}}]{Huang2020ApJ...895..114H}
{Huang}, J., {Luo}, B., {Du}, P., {et~al.} 2020, \apj, 895, 114, \dodoi{10.3847/1538-4357/ab9019}

\bibitem[{{Huang} {et~al.}(2024){Huang}, {Cao}, {Chen}, {Liu}, {Wang}, {You}, \& {Qi}}]{2024icrc.confE1080H}
{Huang}, T.~Q., {Cao}, Z., {Chen}, M., {et~al.} 2024, in 38th International Cosmic Ray Conference, 1080

\bibitem[{{IceCube Collaboration} {et~al.}(2018{\natexlab{a}}){IceCube Collaboration}, {Aartsen}, {Ackermann}, {Adams}, {Aguilar}, {Ahlers}, {Ahrens}, {Al Samarai}, {Altmann}, {Andeen}, {Anderson}, {Ansseau}, {Anton}, {Arg{\"u}elles}, {Auffenberg}, {Axani}, {Bagherpour}, {Bai}, {Barron}, {Barwick}, {Baum}, {Bay}, {Beatty}, {Becker Tjus}, {Becker}, {BenZvi}, {Berley}, {Bernardini}, {Besson}, {Binder}, {Bindig}, {Blaufuss}, {Blot}, {Bohm}, {B{\"o}rner}, {Bos}, {B{\"o}ser}, {Botner}, {Bourbeau}, {Bourbeau}, {Bradascio}, {Braun}, {Brenzke}, {Bretz}, {Bron}, {Brostean-Kaiser}, {Burgman}, {Busse}, {Carver}, {Cheung}, {Chirkin}, {Christov}, {Clark}, {Classen}, {Coenders}, {Collin}, {Conrad}, {Coppin}, {Correa}, {Cowen}, {Cross}, {Dave}, {Day}, {de Andr{\'e}}, {De Clercq}, {DeLaunay}, {Dembinski}, {De Ridder}, {Desiati}, {de Vries}, {de Wasseige}, {de With}, {DeYoung}, {D{\'\i}az-V{\'e}lez}, {di Lorenzo}, {Dujmovic}, {Dumm}, {Dunkman}, {Dvorak}, {Eberhardt}, {Ehrhardt}, {Eichmann}, {Eller}, {Evenson}, {Fahey}, {Fazely}, {Felde}, {Filimonov}, {Finley}, {Flis}, {Franckowiak}, {Friedman}, {Fritz}, {Gaisser}, {Gallagher}, {Gerhardt}, {Ghorbani}, {Glauch}, {Gl{\"u}senkamp}, {Goldschmidt}, {Gonzalez}, {Grant}, {Griffith}, {Haack}, {Hallgren}, {Halzen}, {Hanson}, {Hebecker}, {Heereman}, {Helbing}, {Hellauer}, {Hickford}, {Hignight}, {Hill}, {Hoffman}, {Hoffmann}, {Hoinka}, {Hokanson-Fasig}, {Hoshina}, {Huang}, {Huber}, {Hultqvist}, {H{\"u}nnefeld}, {Hussain}, {In}, {Iovine}, {Ishihara}, {Jacobi}, {Japaridze}, {Jeong}, {Jero}, {Jones}, {Kalaczynski}, {Kang}, {Kappes}, {Kappesser}, {Karg}, {Karle}, {Katz}, {Kauer}, {Keivani}, {Kelley}, {Kheirandish}, {Kim}, {Kim}, {Kintscher}, {Kiryluk}, {Kittler}, {Klein}, {Koirala}, {Kolanoski}, {K{\"o}pke}, {Kopper}, {Kopper}, {Koschinsky}, {Koskinen}, {Kowalski}, {Krings}, {Kroll}, {Kr{\"u}ckl}, {Kunwar}, {Kurahashi}, {Kuwabara}, {Kyriacou}, {Labare}, {Lanfranchi}, {Larson}, {Lauber}, {Leonard}, {Lesiak-Bzdak}, {Leuermann}, {Liu}, {Lozano Mariscal}, {Lu}, {L{\"u}nemann}, {Luszczak}, {Madsen}, {Maggi}, {Mahn}, {Mancina}, {Maruyama}, {Mase}, {Maunu}, {Meagher}, {Medici}, {Meier}, {Menne}, {Merino}, {Meures}, {Miarecki}, {Micallef}, {Moment{\'e}}, {Montaruli}, {Moore}, {Morse}, {Moulai}, {Nahnhauer}, {Nakarmi}, {Naumann}, {Neer}, {Niederhausen}, {Nowicki}, {Nygren}, {Obertacke Pollmann}, {Olivas}, {O'Murchadha}, {O'Sullivan}, {Palczewski}, {Pandya}, {Pankova}, {Peiffer}, {Pepper}, {P{\'e}rez de los Heros}, {Pieloth}, {Pinat}, {Plum}, {Price}, {Przybylski}, {Raab}, {R{\"a}del}, {Rameez}, {Rauch}, {Rawlins}, {Rea}, {Reimann}, {Relethford}, {Relich}, {Resconi}, {Rhode}, {Richman}, {Robertson}, {Rongen}, {Rott}, {Ruhe}, {Ryckbosch}, {Rysewyk}, {Safa}, {S{\"a}lzer}, {Sanchez Herrera}, {Sandrock}, {Sandroos}, {Santander}, {Sarkar}, {Sarkar}, {Satalecka}, {Schlunder}, {Schmidt}, {Schneider}, {Schoenen}, {Sch{\"o}neberg}, {Schumacher}, {Sclafani}, {Seckel}, {Seunarine}, {Soedingrekso}, {Soldin}, {Song}, {Spiczak}, {Spiering}, {Stachurska}, {Stamatikos}, {Stanev}, {Stasik}, {Stein}, {Stettner}, {Steuer}, {Stezelberger}, {Stokstad}, {St{\"o}{\ss}l}, {Strotjohann}, {Stuttard}, {Sullivan}, {Sutherland}, {Taboada}, {Tatar}, {Tenholt}, {Ter-Antonyan}, {Terliuk}, {Tilav}, {Toale}, {Tobin}, {Toennis}, {Toscano}, {Tosi}, {Tselengidou}, {Tung}, {Turcati}, {Turley}, {Ty}, {Unger}, {Usner}, {Vandenbroucke}, {Van Driessche}, {van Eijk}, {van Eijndhoven}, {Vanheule}, {van Santen}, {Vogel}, {Vraeghe}, {Walck}, {Wallace}, {Wallraff}, {Wandler}, {Wandkowsky}, {Waza}, {Weaver}, {Weiss}, {Wendt}, {Werthebach}, {Westerhoff}, {Whelan}, {Whitehorn}, {Wiebe}, {Wiebusch}, {Wille}, {Williams}, {Wills}, {Wolf}, {Wood}, {Wood}, {Woschnagg}, {Xu}, {Xu}, {Xu}, {Yanez}, {Yodh}, {Yoshida}, {Yuan}, {Fermi-LAT Collaboration}, {Abdollahi}, {Ajello}, {Angioni}, {Baldini}, {Ballet}, {Barbiellini}, {Bastieri}, {Bechtol}, {Bellazzini}, {Berenji}, {Bissaldi}, {Blandford}, {Bonino}, {Bottacini}, {Bregeon}, {Bruel}, {Buehler}, {Burnett}, {Burns}, {Buson}, {Cameron}, {Caputo}, {Caraveo}, {Cavazzuti}, {Charles}, {Chen}, {Cheung}, {Chiang}, {Chiaro}, {Ciprini}, {Cohen-Tanugi}, {Conrad}, {Costantin}, {Cutini}, {D'Ammando}, {de Palma}, {Digel}, {Di Lalla}, {Di Mauro}, {Di Venere}, {Dom{\'\i}nguez}, {Favuzzi}, {Franckowiak}, {Fukazawa}, {Funk}, {Fusco}, {Gargano}, {Gasparrini}, {Giglietto}, {Giomi}, {Giommi}, {Giordano}, {Giroletti}, {Glanzman}, {Green}, {Grenier}, {Grondin}, {Guiriec}, {Harding}, {Hayashida}, {Hays}, {Hewitt}, {Horan}, {J{\'o}hannesson}, {Kadler}, {Kensei}, {Kocevski}, {Krauss}, {Kreter}, {Kuss}, {La Mura}, {Larsson}, {Latronico}, {Lemoine-Goumard}, {Li}, {Longo}, {Loparco}, {Lovellette}, {Lubrano}, {Magill}, {Maldera}, {Malyshev}, {Manfreda}, {Mazziotta}, {McEnery}, {Meyer}, {Michelson}, {Mizuno}, {Monzani}, {Morselli}, {Moskalenko}, {Negro}, {Nuss}, {Ojha}, {Omodei}, {Orienti}, {Orlando}, {Palatiello}, {Paliya}, {Perkins}, {Persic}, {Pesce-Rollins}, {Piron}, {Porter}, {Principe}, {Rain{\`o}}, {Rando}, {Rani}, {Razzano}, {Razzaque}, {Reimer}, {Reimer}, {Renault-Tinacci}, {Ritz}, {Rochester}, {Saz Parkinson}, {Sgr{\`o}}, {Siskind}, {Spandre}, {Spinelli}, {Suson}, {Tajima}, {Takahashi}, {Tanaka}, {Thayer}, {Thompson}, {Tibaldo}, {Torres}, {Torresi}, {Tosti}, {Troja}, {Valverde}, {Vianello}, {Vogel}, {Wood}, {Wood}, {Zaharijas}, {MAGIC Collaboration}, {Ahnen}, {Ansoldi}, {Antonelli}, {Arcaro}, {Baack}, {Babi{\'c}}, {Banerjee}, {Bangale}, {Barres de Almeida}, {Barrio}, {Becerra Gonz{\'a}lez}, {Bednarek}, {Bernardini}, {Berti}, {Bhattacharyya}, {Biland}, {Blanch}, {Bonnoli}, {Carosi}, {Carosi}, {Ceribella}, {Chatterjee}, {Colak}, {Colin}, {Colombo}, {Contreras}, {Cortina}, {Covino}, {Cumani}, {Da Vela}, {Dazzi}, {De Angelis}, {De Lotto}, {Delfino}, {Delgado}, {Di Pierro}, {Dom{\'\i}nguez}, {Dominis Prester}, {Dorner}, {Doro}, {Einecke}, {Elsaesser}, {Fallah Ramazani}, {Fern{\'a}ndez-Barral}, {Fidalgo}, {Foffano}, {Pfrang}, {Fonseca}, {Font}, {Franceschini}, {Fruck}, {Galindo}, {Gallozzi}, {Garc{\'\i}a L{\'o}pez}, {Garczarczyk}, {Gaug}, {Giammaria}, {Godinovi{\'c}}, {Gora}, {Guberman}, {Hadasch}, {Hahn}, {Hassan}, {Hayashida}, {Herrera}, {Hose}, {Hrupec}, {Inoue}, {Ishio}, {Konno}, {Kubo}, {Kushida}, {Lelas}, {Lindfors}, {Lombardi}, {Longo}, {L{\'o}pez}, {Maggio}, {Majumdar}, {Makariev}, {Maneva}, {Manganaro}, {Mannheim}, {Maraschi}, {Mariotti}, {Mart{\'\i}nez}, {Masuda}, {Mazin}, {Minev}, {M}, {Mirzoyan}, {Moralejo}, {Moreno}, {Moretti}, {Nagayoshi}, {Neustroev}, {Niedzwiecki}, {Nievas Rosillo}, {Nigro}, {Nilsson}, {Ninci}, {Nishijima}, {Noda}, {Nogu{\'e}s}, {Paiano}, {Palacio}, {Paneque}, {Paoletti}, {Paredes}, {Pedaletti}, {Peresano}, {Persic}, {Prada Moroni}, {Prandini}, {Puljak}, {Rodriguez Garcia}, {Reichardt}, {Rhode}, {Rib{\'o}}, {Rico}, {Righi}, {Rugliancich}, {Saito}, {Satalecka}, {Schweizer}, {Sitarek}, {{\v{S}}nidaric {\textasciiacute}}, {Sobczynska}, {Stamerra}, {Strzys}, {Suri{\'c}}, {Takahashi}, {Tavecchio}, {Temnikov}, {Terzi{\'c}}, {Teshima}, {Torres-Alb{\`a}}, {Treves}, {Tsujimoto}, {Vanzo}, {Vazquez Acosta}, {Vovk}, {Ward}, {Will}, {S}, {Zaric {\textasciiacute}}, {AGILE Team}, {Lucarelli}, {Tavani}, {Piano}, {Donnarumma}, {Pittori}, {Verrecchia}, {Barbiellini}, {Bulgarelli}, {Caraveo}, {Cattaneo}, {Colafrancesco}, {Costa}, {Di Cocco}, {Ferrari}, {Gianotti}, {Giuliani}, {Lipari}, {Mereghetti}, {Morselli}, {Pacciani}, {Paoletti}, {Parmiggiani}, {Pellizzoni}, {Picozza}, {Pilia}, {Rappoldi}, {Trois}, {Vercellone}, {Vittorini}, {ASAS-SN Team}, {Stanek}, {Franckowiak}, {Kochanek}, {Beacom}, {Thompson}, {Holoien}, {Dong}, {Prieto}, {Shappee}, {Holmbo}, {HAWC Collaboration}, {Abeysekara}, {Albert}, {Alfaro}, {Alvarez}, {Arceo}, {Arteaga-Vel{\'a}zquez}, {Avila Rojas}, {Ayala Solares}, {Becerril}, {Belmont-Moreno}, {Bernal}, {Caballero-Mora}, {Capistr{\'a}n}, {Carrami{\~n}ana}, {Casanova}, {Castillo}, {Cotti}, {Cotzomi}, {Couti{\~n}o de Le{\'o}n}, {De Le{\'o}n}, {De la Fuente}, {Diaz Hernandez}, {Dichiara}, {Dingus}, {DuVernois}, {D{\'\i}az-V{\'e}lez}, {Ellsworth}, {Engel}, {Fiorino}, {Fleischhack}, {Fraija}, {Garc{\'\i}a-Gonz{\'a}lez}, {Garfias}, {Gonz{\'a}lez Mu{\~n}oz}, {Gonz{\'a}lez}, {Goodman}, {Hampel-Arias}, {Harding}, {Hernandez}, {Hona}, {Hueyotl-Zahuantitla}, {Hui}, {H{\"u}ntemeyer}, {Iriarte}, {Jardin-Blicq}, {Joshi}, {Kaufmann}, {Kunde}, {Lara}, {Lauer}, {Lee}, {Lennarz}, {Le{\'o}n Vargas}, {Linnemann}, {Longinotti}, {Luis-Raya}, {Luna-Garc{\'\i}a}, {Malone}, {Marinelli}, {Martinez}, {Martinez-Castellanos}, {Mart{\'\i}nez-Castro}, {Mart{\'\i}nez-Huerta}, {Matthews}, {Miranda-Romagnoli}, {Moreno}, {Mostaf{\'a}}, {Nayerhoda}, {Nellen}, {Newbold}, {Nisa}, {Noriega-Papaqui}, {Pelayo}, {Pretz}, {P{\'e}rez-P{\'e}rez}, {Ren}, {Rho}, {Rivi{\`e}re}, {Rosa-Gonz{\'a}lez}, {Rosenberg}, {Ruiz-Velasco}, {Ruiz-Velasco}, {Salesa Greus}, {Sandoval}, {Schneider}, {Schoorlemmer}, {Sinnis}, {Smith}, {Springer}, {Surajbali}, {Tibolla}, {Tollefson}, {Torres}, {Villase{\~n}or}, {Weisgarber}, {Werner}, {Yapici}, {Gaurang}, {Zepeda}, {Zhou}, {{\'A}lvarez}, {H.~E.~S.~S. Collaboration}, {Abdalla}, {Ang{\"u}ner}, {Armand}, {Backes}, {Becherini}, {Berge}, {B{\"o}ttcher}, {Boisson}, {Bolmont}, {Bonnefoy}, {Bordas}, {Brun}, {B{\"u}chele}, {Bulik}, {Caroff}, {Carosi}, {Casanova}, {Cerruti}, {Chakraborty}, {Chandra}, {Chen}, {Colafrancesco}, {Davids}, {Deil}, {Devin}, {Djannati-Ata{\"\i}}, {Egberts}, {Emery}, {Eschbach}, {Fiasson}, {Fontaine}, {Funk}, {F{\"u}{\ss}ling}, {Gallant}, {Gat{\'e}}, {Giavitto}, {Glawion}, {Glicenstein}, {Gottschall}, {Grondin}, {Haupt}, {Henri}, {Hinton}, {Hoischen}, {Holch}, {Huber}, {Jamrozy}, {Jankowsky}, {Jankowsky}, {Jouvin}, {Jung-Richardt}, {Kerszberg}, {Kh{\'e}lifi}, {King}, {Klepser}, {Kluz {\textasciiacute}niak}, {Komin}, {Kraus}, {Lefaucheur}, {Lemi{\`e}re}, {Lemoine-Goumard}, {Lenain}, {Leser}, {Lohse}, {L{\'o}pez-Coto}, {Lorentz}, {Lypova}, {Marandon}, {Guillem Mart{\'\i}-Devesa}, {Maurin}, {Mitchell}, {Moderski}, {Mohamed}, {Mohrmann}, {Moulin}, {Murach}, {de Naurois}, {Niederwanger}, {Niemiec}, {Oakes}, {O'Brien}, {Ohm}, {Ostrowski}, {Oya}, {Panter}, {Parsons}, {Perennes}, {Piel}, {Pita}, {Poireau}, {Priyana Noel}, {Prokoph}, {P{\"u}hlhofer}, {Quirrenbach}, {Raab}, {Rauth}, {Renaud}, {Rieger}, {Rinchiuso}, {Romoli}, {Rowell}, {Rudak},
  {Sasaki}, {Sanchez}, {Schlickeiser}, {Sch{\"u}ssler}, {Schulz}, {Schwanke}, {Seglar-Arroyo}, {Shafi}, {Simoni}, {Sol}, {Stegmann}, {Steppa}, {Tavernier}, {Taylor}, {Tiziani}, {Trichard}, {Tsirou}, {van Eldik}, {van Rensburg}, {van Soelen}, {Veh}, {Vincent}, {Voisin}, {Wagner}, {Wagner}, {Wierzcholska}, {Zanin}, {Zdziarski}, {Zech}, {Ziegler}, {Zorn}, {{\.Z}ywucka}, {INTEGRAL Team}, {Savchenko}, {Ferrigno}, {Bazzano}, {Diehl}, {Kuulkers}, {Laurent}, {Mereghetti}, {Natalucci}, {Panessa}, {Rodi}, {Ubertini}, {Kanata}, Teams, {Morokuma}, {Ohta}, {Tanaka}, {Mori}, {Yamanaka}, {Kawabata}, {Utsumi}, {Nakaoka}, {Kawabata}, {Nagashima}, {Yoshida}, {Matsuoka}, {Itoh}, {Kapteyn Team}, {Keel}, {Liverpool Telescope Team}, {Copperwheat}, {Steele}, {Swift/NuSTAR Team}, {Cenko}, {Cowen}, {DeLaunay}, {Evans}, {Fox}, {Keivani}, {Kennea}, {Marshall}, {Osborne}, {Santander}, {Tohuvavohu}, {Turley}, {VERITAS Collaboration}, {Abeysekara}, {Archer}, {Benbow}, {Bird}, {Brill}, {Brose}, {Buchovecky}, {Buckley}, {Bugaev}, {Christiansen}, {Connolly}, {Cui}, {Daniel}, {Errando}, {Falcone}, {Feng}, {Finley}, {Fortson}, {Furniss}, {Gueta}, {H{\"u}tten}, {Hervet}, {Hughes}, {Humensky}, {Johnson}, {Kaaret}, {Kar}, {Kelley-Hoskins}, {Kertzman}, {Kieda}, {Krause}, {Krennrich}, {Kumar}, {Lang}, {Lin}, {Maier}, {McArthur}, {Moriarty}, {Mukherjee}, {Nieto}, {O'Brien}, {Ong}, {Otte}, {Park}, {Petrashyk}, {Pohl}, {Popkow}, {Pueschel}, {Quinn}, {Ragan}, {Reynolds}, {Richards}, {Roache}, {Rulten}, {Sadeh}, {Santander}, {Scott}, {Sembroski}, {Shahinyan}, {Sushch}, {Tr{\'e}panier}, {Tyler}, {Vassiliev}, {Wakely}, {Weinstein}, {Wells}, {Wilcox}, {Wilhelm}, {Williams}, {Zitzer}, {VLA/B Team}, {Tetarenko}, {Kimball}, {Miller-Jones}, \& {Sivakoff}}]{Aartsen2018Sci...361.1378I}
{IceCube Collaboration}, {Aartsen}, M.~G., {Ackermann}, M., {et~al.} 2018{\natexlab{a}}, Science, 361, eaat1378, \dodoi{10.1126/science.aat1378}

\bibitem[{{IceCube Collaboration} {et~al.}(2018{\natexlab{b}}){IceCube Collaboration}, {Aartsen}, {Ackermann}, {Adams}, {Aguilar}, {Ahlers}, {Ahrens}, {Samarai}, {Altmann}, {Andeen}, {Anderson}, {Ansseau}, {Anton}, {Arg{\"u}elles}, {Arsioli}, {Auffenberg}, {Axani}, {Bagherpour}, {Bai}, {Barron}, {Barwick}, {Baum}, {Bay}, {Beatty}, {Becker Tjus}, {Becker}, {BenZvi}, {Berley}, {Bernardini}, {Besson}, {Binder}, {Bindig}, {Blaufuss}, {Blot}, {Bohm}, {B{\"o}rner}, {Bos}, {B{\"o}ser}, {Botner}, {Bourbeau}, {Bourbeau}, {Bradascio}, {Braun}, {Brenzke}, {Bretz}, {Bron}, {Brostean-Kaiser}, {Burgman}, {Busse}, {Carver}, {Cheung}, {Chirkin}, {Christov}, {Clark}, {Classen}, {Coenders}, {Collin}, {Conrad}, {Coppin}, {Correa}, {Cowen}, {Cross}, {Dave}, {Day}, {de Andr{\'e}}, {De Clercq}, {DeLaunay}, {Dembinski}, {DeRidder}, {Desiati}, {de Vries}, {de Wasseige}, {de With}, {DeYoung}, {D{\'\i}az-V{\'e}lez}, {di Lorenzo}, {Dujmovic}, {Dumm}, {Dunkman}, {Dvorak}, {Eberhardt}, {Ehrhardt}, {Eichmann}, {Eller}, {Evenson}, {Fahey}, {Fazely}, {Felde}, {Filimonov}, {Finley}, {Flis}, {Franckowiak}, {Friedman}, {Fritz}, {Gaisser}, {Gallagher}, {Gerhardt}, {Ghorbani}, {Giommi}, {Glauch}, {Gl{\"u}senkamp}, {Goldschmidt}, {Gonzalez}, {Grant}, {Griffith}, {Haack}, {Hallgren}, {Halzen}, {Hanson}, {Hebecker}, {Heereman}, {Helbing}, {Hellauer}, {Hickford}, {Hignight}, {Hill}, {Hoffman}, {Hoffmann}, {Hoinka}, {Hokanson-Fasig}, {Hoshina}, {Huang}, {Huber}, {Hultqvist}, {H{\"u}nnefeld}, {Hussain}, {In}, {Iovine}, {Ishihara}, {Jacobi}, {Japaridze}, {Jeong}, {Jero}, {Jones}, {Kalaczynski}, {Kang}, {Kappes}, {Kappesser}, {Karg}, {Karle}, {Katz}, {Kauer}, {Keivani}, {Kelley}, {Kheirandish}, {Kim}, {Kim}, {Kintscher}, {Kiryluk}, {Kittler}, {Klein}, {Koirala}, {Kolanoski}, {K{\"o}pke}, {Kopper}, {Kopper}, {Koschinsky}, {Koskinen}, {Kowalski}, {Krammer}, {Krings}, {Kroll}, {Kr{\"u}ckl}, {Kunwar}, {Kurahashi}, {Kuwabara}, {Kyriacou}, {Labare}, {Lanfranchi}, {Larson}, {Lauber}, {Leonard}, {Lesiak-Bzdak}, {Leuermann}, {Liu}, {Lozano Mariscal}, {Lu}, {L{\"u}nemann}, {Luszczak}, {Madsen}, {Maggi}, {Mahn}, {Mancina}, {Maruyama}, {Mase}, {Maunu}, {Meagher}, {Medici}, {Meier}, {Menne}, {Merino}, {Meures}, {Miarecki}, {Micallef}, {Moment{\'e}}, {Montaruli}, {Moore}, {Morse}, {Moulai}, {Nahnhauer}, {Nakarmi}, {Naumann}, {Neer}, {Niederhausen}, {Nowicki}, {Nygren}, {Obertacke Pollmann}, {Olivas}, {O'Murchadha}, {O'Sullivan}, {Padovani}, {Palczewski}, {Pandya}, {Pankova}, {Peiffer}, {Pepper}, {P{\'e}rez de los Heros}, {Pieloth}, {Pinat}, {Plum}, {Price}, {Przybylski}, {Raab}, {R{\"a}del}, {Rameez}, {Rawlins}, {Rea}, {Reimann}, {Relethford}, {Relich}, {Resconi}, {Rhode}, {Richman}, {Robertson}, {Rongen}, {Rott}, {Ruhe}, {Ryckbosch}, {Rysewyk}, {Safa}, {Sahakyan}, {S{\"a}lzer}, {Sanchez Herrera}, {Sandrock}, {Sandroos}, {Santander}, {Sarkar}, {Sarkar}, {Satalecka}, {Schlunder}, {Schmidt}, {Schneider}, {Schoenen}, {Sch{\"o}neberg}, {Schumacher}, {Sclafani}, {Seckel}, {Seunarine}, {Soedingrekso}, {Soldin}, {Song}, {Spiczak}, {Spiering}, {Stachurska}, {Stamatikos}, {Stanev}, {Stasik}, {Stettner}, {Steuer}, {Stezelberger}, {Stokstad}, {St{\"o}{\ss}l}, {Strotjohann}, {Stuttard}, {Sullivan}, {Sutherland}, {Taboada}, {Tatar}, {Tenholt}, {Ter-Antonyan}, {Terliuk}, {Tilav}, {Toale}, {Tobin}, {Toennis}, {Toscano}, {Tosi}, {Tselengidou}, {Tung}, {Turcati}, {Turley}, {Ty}, {Unger}, {Usner}, {Vandenbroucke}, {Van Driessche}, {van Eijk}, {van Eijndhoven}, {Vanheule}, {van Santen}, {Vogel}, {Vraeghe}, {Walck}, {Wallace}, {Wallraff}, {Wandler}, {Wandkowsky}, {Waza}, {Weaver}, {Weiss}, {Wendt}, {Werthebach}, {Westerhoff}, {Whelan}, {Whitehorn}, {Wiebe}, {Wiebusch}, {Wille}, {Williams}, {Wills}, {Wolf}, {Wood}, {Wood}, {Woschnagg}, {Xu}, {Xu}, {Xu}, {Yanez}, {Yodh}, {Yoshida}, \& {Yuan}}]{Aartsen2018Sci...361..147I}
---. 2018{\natexlab{b}}, Science, 361, 147, \dodoi{10.1126/science.aat2890}

\bibitem[{{IceCube Collaboration} {et~al.}(2022){IceCube Collaboration}, {Abbasi}, {Ackermann}, {Adams}, {Aguilar}, {Ahlers}, {Ahrens}, {Alameddine}, {Alispach}, {Alves}, {Amin}, {Andeen}, {Anderson}, {Anton}, {Arg{\"u}elles}, {Ashida}, {Axani}, {Bai}, {Balagopal}, {Barbano}, {Barwick}, {Bastian}, {Basu}, {Baur}, {Bay}, {Beatty}, {Becker}, {Becker Tjus}, {Bellenghi}, {Benzvi}, {Berley}, {Bernardini}, {Besson}, {Binder}, {Bindig}, {Blaufuss}, {Blot}, {Boddenberg}, {Bontempo}, {Borowka}, {B{\"o}ser}, {Botner}, {B{\"o}ttcher}, {Bourbeau}, {Bradascio}, {Braun}, {Brinson}, {Bron}, {Brostean-Kaiser}, {Browne}, {Burgman}, {Burley}, {Busse}, {Campana}, {Carnie-Bronca}, {Chen}, {Chen}, {Chirkin}, {Choi}, {Clark}, {Clark}, {Classen}, {Coleman}, {Collin}, {Conrad}, {Coppin}, {Correa}, {Cowen}, {Cross}, {Dappen}, {Dave}, {de Clercq}, {Delaunay}, {Delgado L{\'o}pez}, {Dembinski}, {Deoskar}, {Desai}, {Desiati}, {de Vries}, {de Wasseige}, {de With}, {Deyoung}, {Diaz}, {D{\'\i}az-V{\'e}lez}, {Dittmer}, {Dujmovic}, {Dunkman}, {Duvernois}, {Dvorak}, {Ehrhardt}, {Eller}, {Engel}, {Erpenbeck}, {Evans}, {Evenson}, {Fan}, {Fazely}, {Fedynitch}, {Feigl}, {Fiedlschuster}, {Fienberg}, {Filimonov}, {Finley}, {Fischer}, {Fox}, {Franckowiak}, {Friedman}, {Fritz}, {F{\"u}rst}, {Gaisser}, {Gallagher}, {Ganster}, {Garcia}, {Garrappa}, {Gerhardt}, {Ghadimi}, {Glaser}, {Glauch}, {Gl{\"u}senkamp}, {Goldschmidt}, {Gonzalez}, {Goswami}, {Grant}, {Gr{\'e}goire}, {Griswold}, {G{\"u}nther}, {Gutjahr}, {Haack}, {Hallgren}, {Halliday}, {Halve}, {Halzen}, {Hanson}, {Hardin}, {Harnisch}, {Haungs}, {Hebecker}, {Helbing}, {Henningsen}, {Hettinger}, {Hickford}, {Hignight}, {Hill}, {Hill}, {Hoffman}, {Hoffmann}, {Hokanson-Fasig}, {Hoshina}, {Huang}, {Huber}, {Huber}, {Hultqvist}, {H{\"u}nnefeld}, {Hussain}, {Hymon}, {in}, {Iovine}, {Ishihara}, {Jansson}, {Japaridze}, {Jeong}, {Jin}, {Jones}, {Kang}, {Kang}, {Kang}, {Kappes}, {Kappesser}, {Kardum}, {Karg}, {Karl}, {Karle}, {Katz}, {Kauer}, {Kellermann}, {Kelley}, {Kheirandish}, {Kin}, {Kintscher}, {Kiryluk}, {Klein}, {Koirala}, {Kolanoski}, {Kontrimas}, {K{\"o}pke}, {Kopper}, {Kopper}, {Koskinen}, {Koundal}, {Kovacevich}, {Kowalski}, {Kozynets}, {Kun}, {Kurahashi}, {Lad}, {Lagunas Gualda}, {Lanfranchi}, {Larson}, {Lauber}, {Lazar}, {Lee}, {Leonard}, {Leszczy{\'n}ska}, {Li}, {Lincetto}, {Liu}, {Liubarska}, {Lohfink}, {Lozano Mariscal}, {Lu}, {Lucarelli}, {Ludwig}, {Luszczak}, {Lyu}, {Ma}, {Madsen}, {Mahn}, {Makino}, {Mancina}, {Mari{\c{s}}}, {Martinez-Soler}, {Maruyama}, {Mase}, {McElroy}, {McNally}, {Mead}, {Meagher}, {Mechbal}, {Medina}, {Meier}, {Meighen-Berger}, {Micallef}, {Mockler}, {Montaruli}, {Moore}, {Morse}, {Moulai}, {Naab}, {Nagai}, {Nahnhauer}, {Naumann}, {Necker}, {Nguyen}, {Niederhausen}, {Nisa}, {Nowicki}, {Nygren}, {Obertack}, {Pollmann}, {Oehler}, {Oeyen}, {Olivas}, {O'Sullivan}, {Pandya}, {Pankova}, {Park}, {Parker}, {Paudel}, {Paul}, {P{\'e}rez de Los Heros}, {Peters}, {Peterson}, {Philippen}, {Pieper}, {Pittermann}, {Pizzuto}, {Plum}, {Popovych}, {Porcelli}, {Prado Rodriguez}, {Price}, {Pries}, {Przybylski}, {Rack-Helleis}, {Raissi}, {Rameez}, {Rawlins}, {Rea}, {Rehman}, {Reichherzer}, {Reimann}, {Renzi}, {Resconi}, {Reusch}, {Rhode}, {Richman}, {Riedel}, {Roberts}, {Robertson}, {Roellinghoff}, {Rongen}, {Rott}, {Ruhe}, {Ryckbosch}, {Rysewyk Cantu}, {Safa}, {Saffer}, {Sanchez Herrera}, {Sandrock}, {Sandroos}, {Santander}, {Sarkar}, {Sarkar}, {Satalecka}, {Schaufel}, {Schieler}, {Schindler}, {Schmidt}, {Schneider}, {Schneider}, {Schr{\"o}der}, {Schumacher}, {Schwefer}, {Sclafani}, {Seckel}, {Seunarine}, {Sharma}, {Shefali}, {Silva}, {Skrzypek}, {Smithers}, {Snihur}, {Soedingrekso}, {Soldin}, {Spannfellner}, {Spiczak}, {Spiering}, {Stachurska}, {Stamatikos}, {Stanev}, {Stein}, {Stettner}, {Steuer}, {Stezelberger}, {Stokstad}, {St{\"u}rwald}, {Stuttard}, {Sullivan}, {Taboada}, {Ter-Antonyan}, {Tilav}, {Tischbein}, {Tollefson}, {T{\"o}nnis}, {Toscano}, {Tosi}, {Trettin}, {Tselengidou}, {Tung}, {Turcati}, {Turcotte}, {Turley}, {Twagirayezu}, {Ty}, {Unland Elorrieta}, {Valtonen-Mattila}, {Vandenbroucke}, {van Eijndhoven}, {Vannerom}, {van Santen}, {Verpoest}, {Walck}, {Watson}, {Weaver}, {Weigel}, {Weindl}, {Weiss}, {Weldert}, {Wendt}, {Werthebach}, {Weyrauch}, {Whitehorn}, {Wiebusch}, {Williams}, {Wolf}, {Woschnagg}, {Wrede}, {Wulff}, {Xu}, {Yanez}, {Yoshida}, {Yu}, {Yuan}, {Zhangan}, \& {Zhelnin}}]{2022Sci...378..538I}
{IceCube Collaboration}, {Abbasi}, R., {Ackermann}, M., {et~al.} 2022, Science, 378, 538, \dodoi{10.1126/science.abg3395}

\bibitem[{{IceCube-Gen2 Collaboration}(2023)}]{Lu2023APS..APRD13008L}
{IceCube-Gen2 Collaboration}. 2023, in APS Meeting Abstracts, Vol. 2023, APS April Meeting Abstracts, D13.008

\bibitem[{{Igumenshchev} {et~al.}(2003){Igumenshchev}, {Narayan}, \& {Abramowicz}}]{Igumenshchev2003ApJ...592.1042I}
{Igumenshchev}, I.~V., {Narayan}, R., \& {Abramowicz}, M.~A. 2003, \apj, 592, 1042, \dodoi{10.1086/375769}

\bibitem[{{Inoue} {et~al.}(2020){Inoue}, {Khangulyan}, \& {Doi}}]{2020ApJ...891L..33I}
{Inoue}, Y., {Khangulyan}, D., \& {Doi}, A. 2020, \apjl, 891, L33, \dodoi{10.3847/2041-8213/ab7661}

\bibitem[{{Keivani} {et~al.}(2018){Keivani}, {Murase}, {Petropoulou}, {Fox}, {Cenko}, {Chaty}, {Coleiro}, {DeLaunay}, {Dimitrakoudis}, {Evans}, {Kennea}, {Marshall}, {Mastichiadis}, {Osborne}, {Santander}, {Tohuvavohu}, \& {Turley}}]{Keivani2018}
{Keivani}, A., {Murase}, K., {Petropoulou}, M., {et~al.} 2018, \apj, 864, 84, \dodoi{10.3847/1538-4357/aad59a}

\bibitem[{{Kelner} {et~al.}(2006){Kelner}, {Aharonian}, \& {Bugayov}}]{Kelner2006PhRvD..74c4018K}
{Kelner}, S.~R., {Aharonian}, F.~A., \& {Bugayov}, V.~V. 2006, \prd, 74, 034018, \dodoi{10.1103/PhysRevD.74.034018}

\bibitem[{{Kheirandish} {et~al.}(2021){Kheirandish}, {Murase}, \& {Kimura}}]{Kheirandish2021ApJ...922...45K}
{Kheirandish}, A., {Murase}, K., \& {Kimura}, S.~S. 2021, \apj, 922, 45, \dodoi{10.3847/1538-4357/ac1c77}

\bibitem[{{Kimura} {et~al.}(2019){Kimura}, {Tomida}, \& {Murase}}]{Kimura2019MNRAS.485..163K}
{Kimura}, S.~S., {Tomida}, K., \& {Murase}, K. 2019, \mnras, 485, 163, \dodoi{10.1093/mnras/stz329}

\bibitem[{{Kun} {et~al.}(2024){Kun}, {Bartos}, {Becker Tjus}, {Biermann}, {Franckowiak}, {Halzen}, {del Palacio}, \& {Woo}}]{Kun2024arXiv240406867K}
{Kun}, E., {Bartos}, I., {Becker Tjus}, J., {et~al.} 2024, arXiv e-prints, arXiv:2404.06867, \dodoi{10.48550/arXiv.2404.06867}

\bibitem[{{Kunz} {et~al.}(2016){Kunz}, {Stone}, \& {Quataert}}]{Kunz2016}
{Kunz}, M.~W., {Stone}, J.~M., \& {Quataert}, E. 2016, \prl, 117, 235101, \dodoi{10.1103/PhysRevLett.117.235101}

\bibitem[{{Liu} {et~al.}(2019){Liu}, {Wang}, {Xue}, {Taylor}, {Wang}, {Li}, \& {Yan}}]{Liu2019}
{Liu}, R.-Y., {Wang}, K., {Xue}, R., {et~al.} 2019, \prd, 99, 063008, \dodoi{10.1103/PhysRevD.99.063008}

\bibitem[{{Liu} {et~al.}(2023){Liu}, {Xue}, {Wang}, {Tan}, \& {B{\"o}ttcher}}]{Liu2023MNRAS.526.5054L}
{Liu}, R.-Y., {Xue}, R., {Wang}, Z.-R., {Tan}, H.-B., \& {B{\"o}ttcher}, M. 2023, \mnras, 526, 5054, \dodoi{10.1093/mnras/stad2911}

\bibitem[{{Lynn} {et~al.}(2014){Lynn}, {Quataert}, {Chandran}, \& {Parrish}}]{Lynn2014ApJ...791...71L}
{Lynn}, J.~W., {Quataert}, E., {Chandran}, B. D.~G., \& {Parrish}, I.~J. 2014, \apj, 791, 71, \dodoi{10.1088/0004-637X/791/1/71}

\bibitem[{{Madejski} \& {Sikora}(2016)}]{Madejski2016ARA&A..54..725M}
{Madejski}, G.~G., \& {Sikora}, M. 2016, \araa, 54, 725, \dodoi{10.1146/annurev-astro-081913-040044}

\bibitem[{{Murase} {et~al.}(2020){Murase}, {Kimura}, \& {M{\'e}sz{\'a}ros}}]{Murase2020PhRvL.125a1101M}
{Murase}, K., {Kimura}, S.~S., \& {M{\'e}sz{\'a}ros}, P. 2020, \prl, 125, 011101, \dodoi{10.1103/PhysRevLett.125.011101}

\bibitem[{Murase {et~al.}(2018)Murase, Oikonomou, \& Petropoulou}]{Murase_2018}
Murase, K., Oikonomou, F., \& Petropoulou, M. 2018, The Astrophysical Journal, 865, 124, \dodoi{10.3847/1538-4357/aada00}

\bibitem[{{Narayan} {et~al.}(2003){Narayan}, {Igumenshchev}, \& {Abramowicz}}]{NIA2003PASJ...55L..69N}
{Narayan}, R., {Igumenshchev}, I.~V., \& {Abramowicz}, M.~A. 2003, \pasj, 55, L69, \dodoi{10.1093/pasj/55.6.L69}

\bibitem[{{Padovani} {et~al.}(2019){Padovani}, {Oikonomou}, {Petropoulou}, {Giommi}, \& {Resconi}}]{Padovani2019MNRAS.484L.104P}
{Padovani}, P., {Oikonomou}, F., {Petropoulou}, M., {Giommi}, P., \& {Resconi}, E. 2019, \mnras, 484, L104, \dodoi{10.1093/mnrasl/slz011}

\bibitem[{{Pringle}(1981)}]{Pringle1981ARA&A..19..137P}
{Pringle}, J.~E. 1981, \araa, 19, 137, \dodoi{10.1146/annurev.aa.19.090181.001033}

\bibitem[{{Ricci} {et~al.}(2018){Ricci}, {Ho}, {Fabian}, {Trakhtenbrot}, {Koss}, {Ueda}, {Lohfink}, {Shimizu}, {Bauer}, {Mushotzky}, {Schawinski}, {Paltani}, {Lamperti}, {Treister}, \& {Oh}}]{Ricci2018MNRAS.480.1819R}
{Ricci}, C., {Ho}, L.~C., {Fabian}, A.~C., {et~al.} 2018, \mnras, 480, 1819, \dodoi{10.1093/mnras/sty1879}

\bibitem[{{Rieger} \& {Aharonian}(2008)}]{Rieger2008}
{Rieger}, F.~M., \& {Aharonian}, F.~A. 2008, \aap, 479, L5, \dodoi{10.1051/0004-6361:20078706}

\bibitem[{{Ripperda} {et~al.}(2020){Ripperda}, {Bacchini}, \& {Philippov}}]{Ripperda2020ApJ...900..100R}
{Ripperda}, B., {Bacchini}, F., \& {Philippov}, A.~A. 2020, \apj, 900, 100, \dodoi{10.3847/1538-4357/ababab}

\bibitem[{{Ripperda} {et~al.}(2022){Ripperda}, {Liska}, {Chatterjee}, {Musoke}, {Philippov}, {Markoff}, {Tchekhovskoy}, \& {Younsi}}]{Ripperda2022}
{Ripperda}, B., {Liska}, M., {Chatterjee}, K., {et~al.} 2022, \apjl, 924, L32, \dodoi{10.3847/2041-8213/ac46a1}

\bibitem[{{Rodrigues} {et~al.}(2019){Rodrigues}, {Gao}, {Fedynitch}, {Palladino}, \& {Winter}}]{Rodrigues2019}
{Rodrigues}, X., {Gao}, S., {Fedynitch}, A., {Palladino}, A., \& {Winter}, W. 2019, \apjl, 874, L29, \dodoi{10.3847/2041-8213/ab1267}

\bibitem[{{Sahakyan}(2018)}]{Sahakyan2018}
{Sahakyan}, N. 2018, \apj, 866, 109, \dodoi{10.3847/1538-4357/aadade}

\bibitem[{{Sironi} \& {Spitkovsky}(2014)}]{Sironi2014ApJ...783L..21S}
{Sironi}, L., \& {Spitkovsky}, A. 2014, \apjl, 783, L21, \dodoi{10.1088/2041-8205/783/1/L21}

\bibitem[{{Tchekhovskoy} {et~al.}(2014){Tchekhovskoy}, {Metzger}, {Giannios}, \& {Kelley}}]{Tchekhovskoy2014MNRAS.437.2744T}
{Tchekhovskoy}, A., {Metzger}, B.~D., {Giannios}, D., \& {Kelley}, L.~Z. 2014, \mnras, 437, 2744, \dodoi{10.1093/mnras/stt2085}

\bibitem[{{Tchekhovskoy} {et~al.}(2011){Tchekhovskoy}, {Narayan}, \& {McKinney}}]{Tchekhovskoy2011MNRAS.418L..79T}
{Tchekhovskoy}, A., {Narayan}, R., \& {McKinney}, J.~C. 2011, \mnras, 418, L79, \dodoi{10.1111/j.1745-3933.2011.01147.x}

\bibitem[{{Trakhtenbrot} {et~al.}(2019){Trakhtenbrot}, {Arcavi}, {Ricci}, {Tacchella}, {Stern}, {Netzer}, {Jonker}, {Horesh}, {Mej{\'\i}a-Restrepo}, {Hosseinzadeh}, {Hallefors}, {Howell}, {McCully}, {Balokovi{\'c}}, {Heida}, {Kamraj}, {Lansbury}, {Wyrzykowski}, {Gromadzki}, {Hamanowicz}, {Cenko}, {Sand}, {Hsiao}, {Phillips}, {Diamond}, {Kara}, {Gendreau}, {Arzoumanian}, \& {Remillard}}]{Trakhtenbrot2019NatAs...3..242T}
{Trakhtenbrot}, B., {Arcavi}, I., {Ricci}, C., {et~al.} 2019, Nature Astronomy, 3, 242, \dodoi{10.1038/s41550-018-0661-3}

\bibitem[{{Wang} {et~al.}(2022){Wang}, {Liu}, {Li}, {Wang}, \& {Dai}}]{Wang2022}
{Wang}, K., {Liu}, R.-Y., {Li}, Z., {Wang}, X.-Y., \& {Dai}, Z.-G. 2022, Universe, 9, 1, \dodoi{10.3390/universe9010001}

\bibitem[{{Werner} {et~al.}(2018){Werner}, {Uzdensky}, {Begelman}, {Cerutti}, \& {Nalewajko}}]{Werner2018MNRAS.473.4840W}
{Werner}, G.~R., {Uzdensky}, D.~A., {Begelman}, M.~C., {Cerutti}, B., \& {Nalewajko}, K. 2018, \mnras, 473, 4840, \dodoi{10.1093/mnras/stx2530}

\bibitem[{{Xue} {et~al.}(2019){Xue}, {Liu}, {Petropoulou}, {Oikonomou}, {Wang}, {Wang}, \& {Wang}}]{Xue2019}
{Xue}, R., {Liu}, R.-Y., {Petropoulou}, M., {et~al.} 2019, \apj, 886, 23, \dodoi{10.3847/1538-4357/ab4b44}

\bibitem[{{Xue} {et~al.}(2021){Xue}, {Liu}, {Wang}, {Ding}, \& {Wang}}]{Xue2021ApJ...906...51X}
{Xue}, R., {Liu}, R.-Y., {Wang}, Z.-R., {Ding}, N., \& {Wang}, X.-Y. 2021, \apj, 906, 51, \dodoi{10.3847/1538-4357/abc886}

\bibitem[{{Yuan} \& {Narayan}(2014)}]{Yuan2014ARA&A..52..529Y}
{Yuan}, F., \& {Narayan}, R. 2014, \araa, 52, 529, \dodoi{10.1146/annurev-astro-082812-141003}

\bibitem[{{Yuan} {et~al.}(2003){Yuan}, {Quataert}, \& {Narayan}}]{Yuan2003}
{Yuan}, F., {Quataert}, E., \& {Narayan}, R. 2003, \apj, 598, 301, \dodoi{10.1086/378716}

\bibitem[{Zathul {et~al.}(2024)Zathul, Moulai, Fang, \& Halzen}]{zathul2024ngc1068informedunderstandingneutrino}
Zathul, A.~K., Moulai, M., Fang, K., \& Halzen, F. 2024, An NGC 1068-Informed Understanding of Neutrino Emission of the Active Galactic Nucleus TXS 0506+056.
\newblock \doarXiv{2411.14598}

\bibitem[{{Zhang} {et~al.}(2020){Zhang}, {Petropoulou}, {Murase}, \& {Oikonomou}}]{ZhangB2020}
{Zhang}, B.~T., {Petropoulou}, M., {Murase}, K., \& {Oikonomou}, F. 2020, \apj, 889, 118, \dodoi{10.3847/1538-4357/ab659a}

\end{thebibliography}
\bibliographystyle{aasjournal}

\end{document}